\documentclass[pre,twocolumn,nobalancelastpage,amsmath,amssymb,showpacs,
nofootinbib, superscriptaddress]{revtex4-1}

\usepackage{graphics}      % standard graphics specifications
\usepackage{graphicx}      % alternative graphics specifications
\usepackage{longtable}     % helps with long table options
\usepackage{url}           % for on-line citations
\usepackage{bm}            % special 'bold-math' package
\usepackage{xcolor}
\usepackage{amsmath}
\usepackage{dcolumn}
\usepackage{mathtools}
\usepackage{hyperref}

\begin{document}
\title{Shear-Driven Flow of Athermal, Frictionless, Spherocylinder Suspensions in Two Dimensions:  Spatial Structure and Correlations}
\author{Theodore A. Marschall}
\affiliation{Department of Physics and Astronomy, University of Rochester, Rochester, NY 14627}
\author{S. Teitel}
\affiliation{Department of Physics and Astronomy, University of Rochester, Rochester, NY 14627}
\date{\today}

\begin{abstract}
We use numerical simulations to study the flow of athermal, frictionless, soft-core two dimensional spherocylinders driven by a uniform steady-state simple shear  applied at a fixed volume and a fixed finite strain rate $\dot\gamma$.  Energy dissipation is via a viscous drag with respect to a uniformly sheared host fluid, giving a simple model for flow in a non-Brownian suspension with Newtonian rheology.  We study the resulting spatial structure of the sheared system, and compute correlation functions of the velocity, the particle density, the nematic order parameter, and the particle angular velocity.  Correlations of density, nematic order, and angular velocity are shown to be short ranged both below and above jamming.
We compare a system of size-bidisperse particles with a system of size-monodisperse particles, and argue how  differences in spatial order as the packing increases leads to differences in the global nematic order parameter.  We consider the effect of shearing on initially well ordered configurations, and show that in many cases the shearing acts to destroy the order, leading to the same steady-state ensemble as found when starting from  random initial configurations.
\end{abstract}
\maketitle

%%%%%%%%%%%%%%%%%%%%%%%%%%%%%%%%%%%%%%%%%%%%%%%%%%%%%%%%%%%%%%%%%%%%%%%%%%%%%
\section{Introduction}
\label{sec:intro}

In a system of athermal granular particles with only repulsive contact interactions, as the packing fraction of particles $\phi$ increases, the system undergoes a jamming transition \cite{OHern,LiuNagel} at a critical $\phi_J$.  For $\phi<\phi_J$ the system behaves similar to a liquid, while for $\phi>\phi_J$ the system behaves like a rigid but disordered solid.  
One way to probe the jamming transition is through the application of a simple shear deformation to the system.  For an infinite system in the ``thermodynamic limit," if one applies a simple shear stress $\sigma$ no matter how small, then if the system is below $\phi_J$  the system responds with a simple shear flow, with a velocity profile that varies linearly in the direction transverse to the flow.  Above $\phi_J$, the application of a small shear stress causes the system to have an elastic shear distortion determined by the finite shear modulus of the solid phase; the system does not flow.  However, if $\sigma$ exceeds a critical yield stress $\sigma_0$, then plastic deformations cause the solid to flow.  The point where this yield stress $\sigma_0(\phi)$ vanishes upon decreasing $\phi$  then determines the shear-driven jamming transition $\phi_J$ \cite{OlssonTeitelPRL,OlssonTeitelPRE,VagbergOlssonTeitel}.  For frictionless particles, such as those considered in this work, $\sigma_0$ vanishes continuously \cite{OlssonTeitelPRL,OlssonTeitelPRE} as $\phi\to\phi_J$ from above.

Many numerical studies of the jamming transition, and granular materials more generally, have used spherically shaped particles for simplicity.  
It is therefore interesting to ask how behavior  is modified if the particles have shapes with a lower rotational symmetry \cite{Borzsonyi.Soft.2013}.  In a recent work \cite{MT1} we considered the simple shear-driven jamming of a suspension of athermal, bidisperse, overdamped, frictionless, spherocylinders in two dimensions (2D), uniformly sheared at a fixed strain rate $\dot\gamma$.  In that work we  considered  the global rheology of the system, investigating how pressure, deviatoric shear stress, and macroscopic friction vary with particle packing fraction $\phi$, shear strain rate $\dot\gamma$, and particle asphericity $\alpha$.  In a subsequent work \cite{MT2} we focused on the rotational motion and nematic orientational ordering of spherocylinders in simple shear flow, arguing for a crossover in behavior as the particle packing fraction increased.  At small packings $\phi$, the particle rotations are single-particle-like, though perturbed by inter-particle collisions.  At larger $\phi$, approaching and going above jamming, the geometry of the dense packings inhibits particle rotations, which become a random Poisson-type process.  This crossover leads to a non-monotonic behavior of the average particle angular velocity $\langle\dot\theta_i\rangle/\dot\gamma$, and the magnitude of the nematic ordering
$S_2$, as $\phi$ increases.  We also argued that nematic orientational ordering was a consequence of the shearing acting like an ordering field, rather than  due to long-range cooperative behavior among the particles.  

In this work we continue our studies of this 2D spherocylinder model, but now concentrating on the spatial structure of the sheared system, and the spatial correlations of various quantities, including the particle density, nematic order parameter, and angular velocity.   We confirm the assertion in \cite{MT2}, that there is no long-range cooperative behavior causing the finite nematic ordering, by showing that correlations of the nematic order parameter are short-ranged.  By comparing the behavior of a size-bidisperse system of particles with a size-monodisperse system, and finding that the monodisperse system has a greater local spatial ordering, we find further evidence for our claim in \cite{MT2} that at large $\phi$ it is the specific geometry of the dense packing that determines particle rotations and nematic ordering.

The rest of this paper is organized as follows.  In Sec.~\ref{sec:modelMethod} we discuss our model and simulation methods.  In Sec.~\ref{sec:Bidisperse} we present our results for a size-bidisperse system of particles.  We consider both the case of moderately elongated spherocylinders with $\alpha=4$, as well as nearly circular spherocylinders with $\alpha=0.01$.  In Sec.~\ref{sec:Monodisperse} we present our results for a size-monodisperse system of particles, considering only the case of elongated particles with $\alpha=4$.  In Sec.~\ref{sec:HOR} we consider what happens when one starts the shearing from an initially well ordered state, as opposed to the random initial states considered in the rest of our work.  We find that in many cases, the sheared steady-state ensemble becomes independent of the initial configuration after sufficiently long shearing.  In Sec.~\ref{sec:discus} we summarize our conclusions.

\section{Model and Simulation Method}
\label{sec:modelMethod}

Our model is intended to describe a system of particles in a suspending host medium, rather than a dry granular material.  Dissipation is taken to be due to a viscous drag between the particles and the host medium, rather than due to inelastic particle collisions, and the resulting rheology in the dilute phase is Newtonian.
As this work is a continuation of our  prior work on this system, the description of the model presented here is abbreviated.  We  refer the reader to our earlier works \cite{MT1,MT2} for a  discussion of the broader context of, and motivation for, our model, a more complete list of references, and more details of the derivation of our equations of motion.

%\subsection{Model}
%\label{sec:Model}

\begin{figure}
\centering
\includegraphics[width=2.5in]{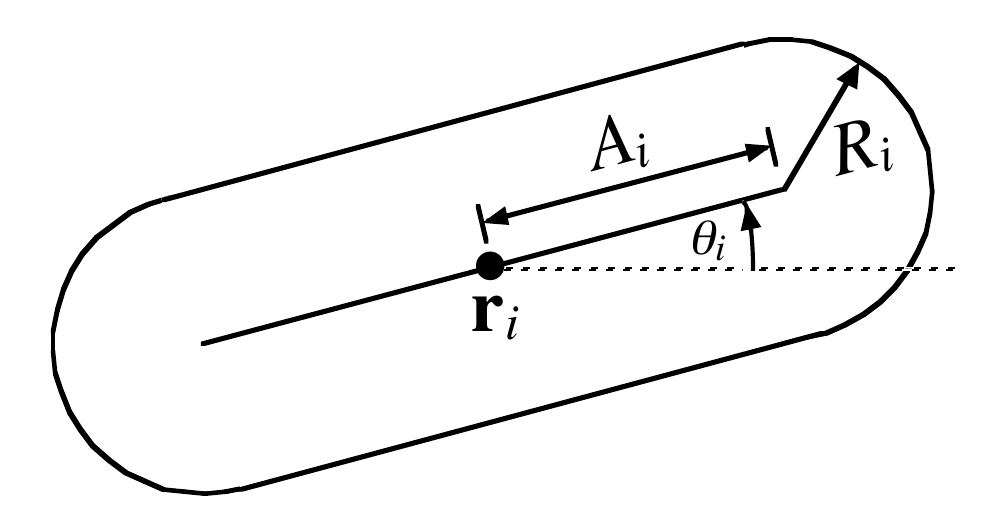}
\caption{ An isolated spherocylinder indicating the spine half-length $A_i$, end cap radius $R_i$, center of mass position $\mathbf{r}_i$, and angle of orientation $\theta_i$. }
\label{sphero} 
\end{figure}

We consider a two dimensional system of $N$,  athermal, frictionless spherocylinders, consisting of a rectangle with two semi-circular end caps, as illustrated in Fig.~\ref{sphero}.  The half-length of the rectangle of particle $i$ is $A_i$, the radius is $R_i$, and we define the asphericity $\alpha_i$ as,
\begin{equation}
\alpha_i=A_i/R_i
\end{equation}
so that $\alpha=0$ is a pure circular particle.  The ``spine" of the spherocylinder is the axis of length $2A_i$ that goes down the center of the rectangle.  For every point on the perimeter of the spherocylinder, the shortest distance from the spine is $R_i$.  The center of mass of the particle is $\mathbf{r}_i$ and the angle $\theta_i$ denotes the orientation of the spine with respect to the flow direction $\mathbf{\hat x}$.  
Our system box has lengths $L_x$ and $L_y$ in the $\mathbf{\hat x}$ and $\mathbf{\hat y}$ directions, respectively.  We will in general take $L_x=L_y\equiv L$ unless otherwise noted.
If $\mathcal{A}_i$ is the area of spherocylinder $i$, the packing fraction $\phi$ is,
\begin{equation}
\phi=\frac{1}{L^2}\sum_{i=1}^N\mathcal{A}_i.
\end{equation}
All  particles in our systems are taken to have an equal asphericity $\alpha$.  In Sec.~\ref{sec:Bidisperse} we will consider a system of particles that are bidisperse in size, with equal numbers of small and big particles with length scales in the ratio $R_b/R_s=1.4$.  In Sec.~\ref{sec:Monodisperse} we will consider a system of particles that are monodisperse in size.

Periodic boundary conditions are taken along $\mathbf{\hat x}$, while Lees-Edward boundary conditions \cite{LeesEdwards} are taken along $\mathbf{\hat y}$ to introduce a simple shear strain $\gamma$.  We take $\gamma =\dot\gamma t$ to model simple shear flow in the $\mathbf{\hat x}$ direction at a fixed finite strain rate $\dot\gamma$.  Particles interact with each other via elastic contact interactions.  
Defining $r_{ij}$ as the shortest distance between the spines of spherocylinders $i$ and $j$ \cite{Pournin.GranulMat.2005}, and $d_{ij}=R_i+R_j$, two spherocylinders are in contact whenever $r_{ij}<d_{ij}$.  In this case there is a repulsive harmonic interaction between the particles, with the force on $i$ given by,
\begin{equation}
\mathbf{F}_{ij}^\mathrm{el}=\frac{k_e}{d_{ij}}\left(1-\frac{r_{ij}}{d_{ij}}\right)\mathbf{\hat n}_{ij},
\end{equation}
where $k_e$ is the particle stiffness and  $\mathbf{\hat n}_{ij}$ the unit vector pointing normally inwards to particle $i$ at the point of contact with $j$.  $\mathbf{F}_{ij}^\mathrm{el}$ acts at the contact point, which is located a distance $(R_i/d_{ij})r_{ij}$ from the spine of particle $i$, along the cord $r_{ij}$, and gives rise to a torque on particle $i$,
\begin{equation}
\boldsymbol{\tau}_{ij}^\mathrm{el}=\mathbf{\hat z} \tau_{ij}^\mathrm{el}=\mathbf{s}_{ij}\times\mathbf{F}_{ij}^\mathrm{el},
\end{equation}
where $\mathbf{s}_{ij}$ is the moment arm from the center of mass of $i$ to its point of contact with $j$.  The total elastic force and torque on particle $i$ are then
\begin{equation}
\mathbf{F}_i^\mathrm{el}=\sum_j \mathbf{F}_{ij}^\mathrm{el},\qquad
\tau_i^\mathrm{el}=\sum_j \tau_{ij}^\mathrm{el}
\end{equation}
where the sums are over all particles $j$ in contact with $i$.

Energy dissipation is due to  a viscous drag between the particles and the affinely sheared host medium.
The viscous drag force density at position $\mathbf{r}$ on particle $i$ is 
\begin{equation}
\mathbf{f}_i^\mathrm{dis}(\mathbf{r})=-k_d[\mathbf{v}_i(\mathbf{r})-\mathbf{v}_\mathrm{host}(\mathbf{r})],
\end{equation}
where $k_d$ is a viscous damping coefficient,  $\mathbf{v}_\mathrm{host}(\mathbf{r})$ is the local velocity of the host medium, which for simple shearing in the $\mathbf{\hat x}$ direction is,
\begin{equation}
\mathbf{v}_\mathrm{host}(\mathbf{r})=\dot\gamma y \mathbf{\hat x}, 
\end{equation}
and
$\mathbf{v}_i(\mathbf{r})$ is the local velocity of the particle at position $\mathbf{r}$,
\begin{equation}
\mathbf{v}_i(\mathbf{r})=\mathbf{\dot r}_i+\dot\theta_i\mathbf{\hat z}\times (\mathbf{r}-\mathbf{r}_i),
\end{equation}
where $\mathbf{\dot r}_i=d\mathbf{r}_i/dt$ is the center of mass velocity of the particle and $\dot\theta_i$ is its angular velocity about the center of mass. 

The total viscous drag force on  particle $i$ is then taken as,
\begin{equation}
\mathbf{F}_i^\mathrm{dis}=\int_i d^2r\,\mathbf{f}_i^\mathrm{dis}(\mathbf{r}),
\end{equation}
where the integral is over the area of particle $i$.
The corresponding dissipative torque is,
\begin{equation}
\boldsymbol{\tau}_i^\mathrm{dis}=\mathbf{\hat z}\tau_i^\mathrm{dis}=\int_i d^2r\,(\mathbf{r}-\mathbf{r}_i)\times \mathbf{f}_i^\mathrm{dis}(\mathbf{r}).
\end{equation}

The above elastic and dissipative forces are the only forces included in our model; there are no inter-particle dissipative or frictional forces.  We will carry out our simulations in the overdampled (low particle mass) limit, where the total force and torque on each particle are damped to zero, 
\begin{equation}
\mathbf{F}_i^{\mathrm{el}} + \mathbf{F}_i^{\mathrm{dis}} = 0, 
\quad
\tau_i^{\mathrm{el}} + \tau_i^{\mathrm{dis}} = 0.
\end{equation} 
The resulting translational and rotational equations of motion for particle $i$ can then be written as \cite{MT1},
\begin{align}
\dot{\mathbf{r}}_i &=    \dot{\gamma}y_i{\mathbf{\hat x}}+\dfrac{\mathbf{F}_i^{\mathrm{el}}}{k_d \mathcal{A}_i},
\label{eq:ri_eom} \\
\dot{\theta}_i &= - \dot{\gamma} f(\theta_i)+ \dfrac{\tau_i^{\mathrm{el}}}{k_d  \mathcal{A}_iI_i},
\label{eq:theta_eom}
\end{align}
where $\mathcal{A}_i$ is the area of particle $i$, $I_i$ is the trace of the particle's moment of inertia tensor, and 
\begin{equation}
f(\theta)=\frac{1}{2}\left[1-\left({\Delta I_i}/{I_i}\right)\cos 2\theta\right],
\label{eftheta}
\end{equation}
where  $\Delta I_i$ is the absolute value of the difference of the two eigenvalues of the moment of inertia tensor.  We assume a uniform constant  mass density for both our small and big particles.

One of the distinguishing features of aspherical particles in simple shear flow is that they tumble as they flow, and that they show a finite nematic orientational ordering $\mathbf{S}_2$  \cite{MT2,MKOT,Campbell,Guo1,Guo2,Borzsonyi1,Borzsonyi2,Wegner,Nath}, with the spines of the spherocylinders tending to align  about a given direction.   The extent of the alignment is given by the magnitude of the nematic order parameter $S_2$, while the direction of alignment is given by the angle $\theta_2$ with respect to the flow direction $\mathbf{\hat x}$.
For a two dimensional system, these can be computed by \cite{Torquato},
\begin{equation}
S_2=\sqrt{\left[\dfrac{1}{N}\sum_{i=1}^N\cos(2\theta_i)\right]^2+\left[\dfrac{1}{N}\sum_{i=1}^N\sin(2\theta_i)\right]^2 },
\label{eS2g1}
\end{equation}
and
\begin{equation}
\tan[2\theta_2]=\left[\dfrac{1}{N}\sum_{i=1}^N\sin(2\theta_i)\right] \bigg/  \left[\dfrac{1}{N}\sum_{i=1}^N\cos(2\theta_i)\right].
\label{eS2g2}
\end{equation}
To compute the nematic order parameter of a specific configuration, the square brackets in the above expressions represent sums over the $N$ particles in the system.  To compute the ensemble averaged nematic order parameter, the square brackets should be taken as both a sum over the $N$ particles in the system, as well as an average over all configurations in the sheared steady state.

For our simulations we  take $2 R_s = 1$ as the unit of distance, $k_e = 1$ as the unit of energy, and $t_0 = (2 R_s)^2 k_d\mathcal{A}_s / k_e = 1$ as the unit of time.  For simplicity we take the viscous drag $k_d$ to vary with particle size so that $k_d\mathcal{A}_i=1$ for all particles.
We numerically integrate the equations of motion (\ref{eq:ri_eom}) and (\ref{eq:theta_eom}) using a two-stage Heun method with a step size of $\Delta t = 0.02$.
Except for the simulations discussed in Sec.~\ref{sec:HOR}, we begin each shearing run in a finite energy configuration at the desired packing fraction $\phi$, with random initial particle positions and orientations.
To generate such initial configurations we place the spherocylinders in the system one-by-one, while rejecting and retrying any time a new placement would lead to an unphysical overlap where the spines of two spherocylinders  intersect.  In general we use $N=1024$ particles.
Our simulations typically extend to total strains of at least $\gamma\approx 150$.  Discarding an   initial $\Delta\gamma\approx 20$ of the strain  from the averaging so as to eliminate transients effects, we find that our steady state averages are generally insensitive to the particular starting configuration.  
Note, we restrict the strain coordinate $\gamma$ used in our Lees-Edwards boundary condition  to the range $\gamma\in \left(-L_x/2L_y, L_x/2L_y\right]$; whenever it exceeds this maximum it is reset by taking $\gamma \to \gamma - L_x/Ly$, allowing us to shear to arbitrarily large total strains.

\section{Size-Bidisperse Particles}
\label{sec:Bidisperse}

\begin{figure}
\includegraphics[width=3.5in]{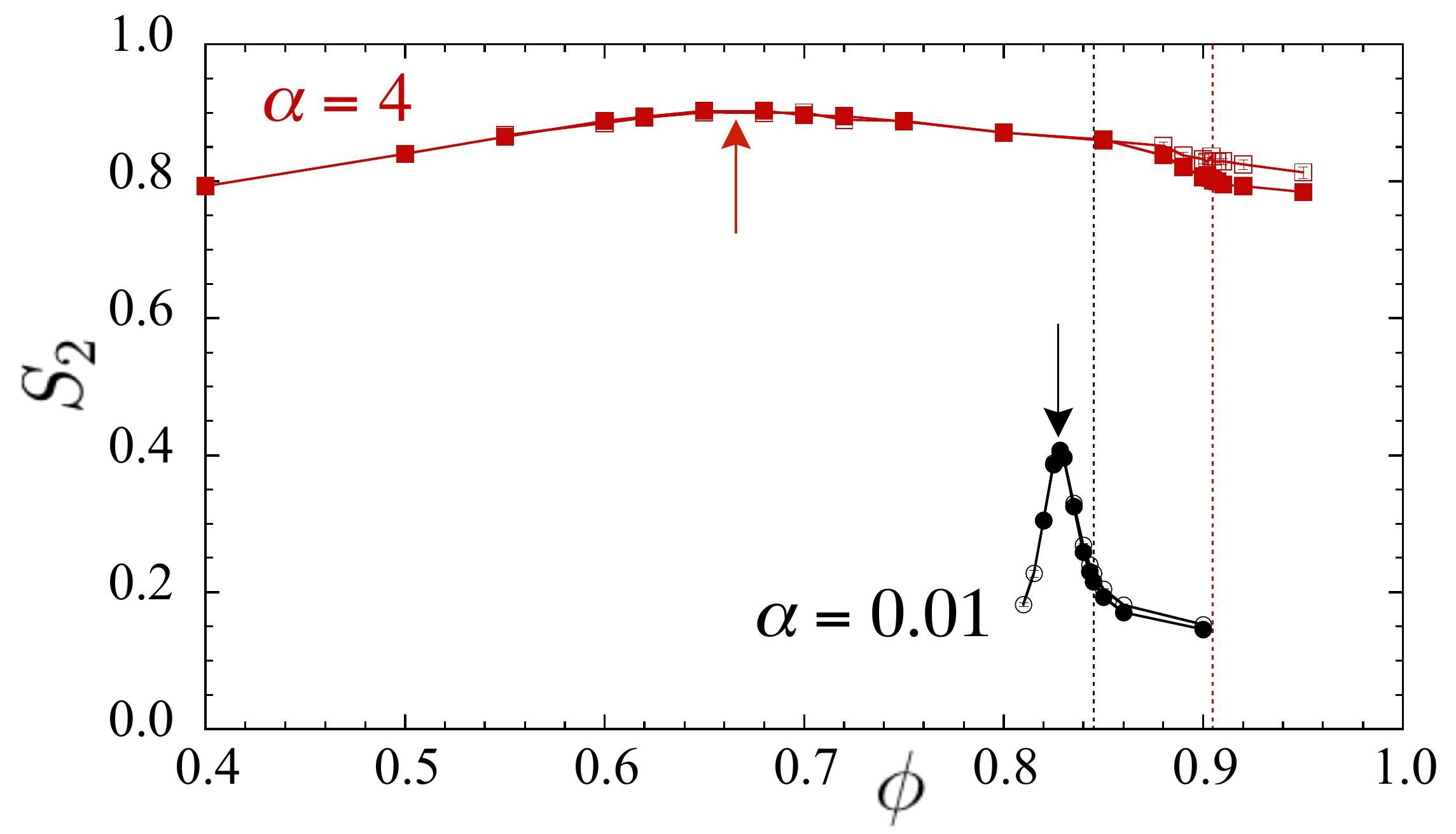}
\caption{
Magnitude of the nematic order parameter $S_2$ vs packing $\phi$, for elongated spherocylinders of $\alpha=4$ and nearly circular spherocylinders of $\alpha=0.01$, in simple shear steady-state.  
Dotted vertical lines locate the respective jamming transitions, $\phi_J(\alpha=4)\approx0.906$ and $\phi_J(\alpha=0.01)\approx0.845$.  
The vertical arrows indicate the location of the maxima in $S_2$ at $\phi_{S_2\,\mathrm{max}}\approx 0.67$ and 0.83 for $\alpha=4$ and 0.01 respectively. 
For each case we show results at two different strain rates.
For $\alpha=4$, solid symbols are at strain rate $\dot\gamma=10^{-5}$, while open symbols are at $\dot\gamma=4\times 10^{-5}$; for $\alpha=0.01$, solid symbols are for $\dot\gamma=4\times 10^{-7}$, while open symbols are for $\dot\gamma=10^{-6}$.  
}
\label{S2-vs-phi} 
\end{figure}

In this section we consider a system of size-bidisperse particles, with equal numbers of big and small spherocylinders with radii in the ratio of $R_b/R_s=1.4$.  We will consider both the case of moderately elongated spherocylinders with $\alpha=4$, and nearly circular spherocylinders with $\alpha=0.01$.  To set the scale for the various packing fractions $\phi$ that we will consider, in Fig.~\ref{S2-vs-phi} we show a plot of the magnitude of the nematic order parameter $S_2$ vs $\phi$ for these two cases.  As noted in our previous work \cite{MT2,MKOT}, $S_2$ is non-monotonic in $\phi$, with a peak at $\phi_{S_2\,\mathrm{max}}$ that lies somewhat below the jamming $\phi_J$.  For $\alpha=4$ we have $\phi_{S_2\,\mathrm{max}}\approx 0.67$ and $\phi_J\approx 0.906$; for $\alpha=0.01$, we have $\phi_{S_2\,\mathrm{max}}\approx 0.83$ and $\phi_J\approx 0.845$.

\begin{figure}
\centering
\includegraphics[width=3.5in]{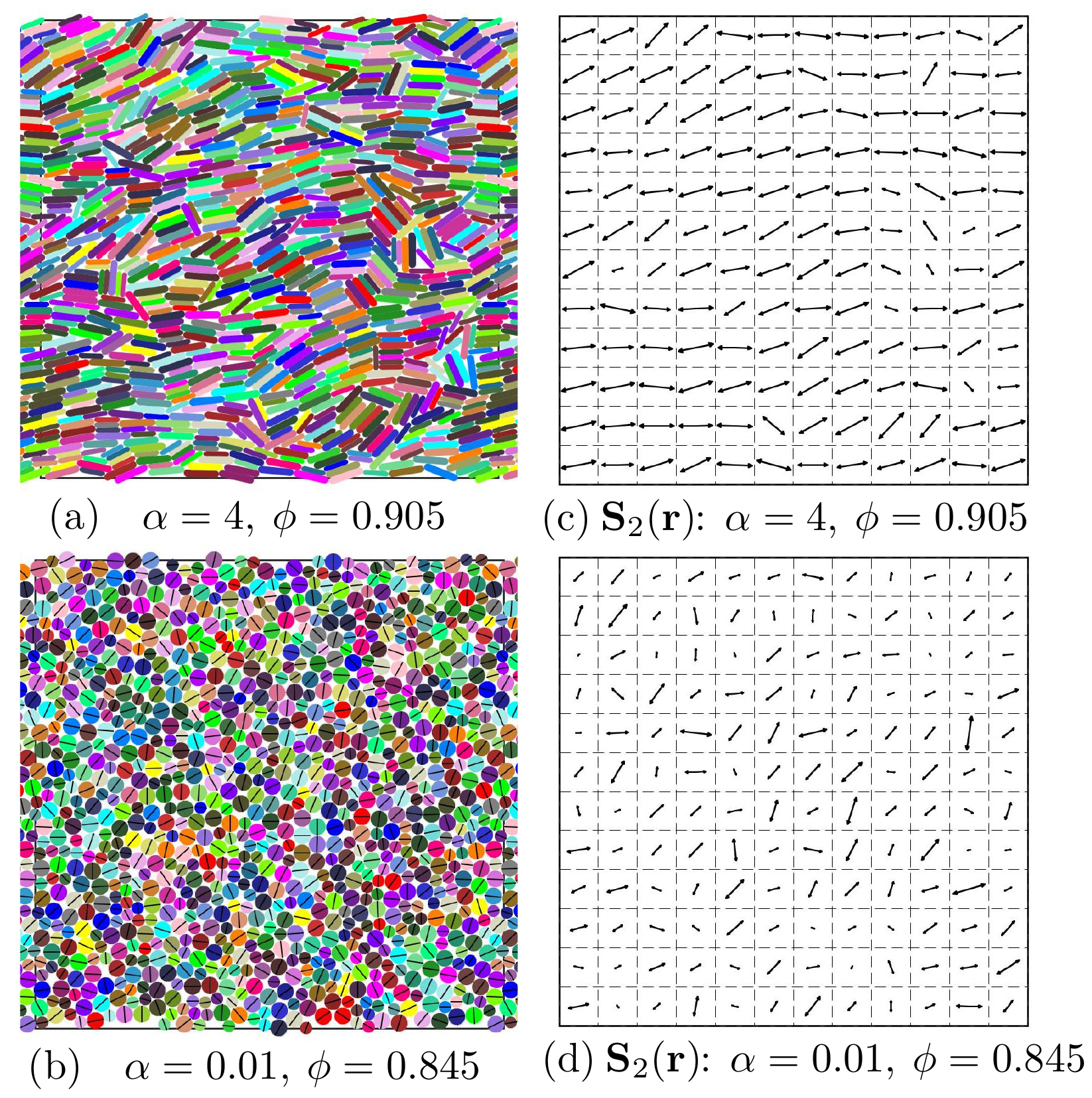}
\caption{Snapshot configurations in simple sheared steady-state with strain rate $\dot\gamma=10^{-6}$ for spherocylinders of asphericity (a)  $\alpha=4$ at packing $\phi=0.905$ near the jamming $\phi_J=0.906$, and (b) $\alpha=0.01$ at packing $\phi_J=0.845$.  In (b) straight lines on particles indicate the directions of the spines.  Different colors are used to help distinguish different particles and have no other meaning. (c) and (d) show the corresponding configurations of the  local nematic order parameter $\mathbf{S}_2(\mathbf{r})$, obtained by averaging over all particles whose center of mass $\mathbf{r}_i$ is contained in each square grid cell.  Corresponding animations, showing the evolutions of these configurations under shearing, are available in our Supplemental Material \cite{SM}.
}
\label{configs} 
\end{figure}

We start with a qualitative description of the spatial structure of the system.
In Fig.~\ref{configs} we show snapshots of typical configurations sampled during steady-state shearing at strain rate $\dot\gamma=10^{-6}$.  In Fig.~\ref{configs}(a) we show a system with $\alpha=4$ at packing $\phi=0.905$, very close to the jamming $\phi_J=0.906$.  In Fig.~\ref{configs}(b) we show a system with $\alpha=0.01$ at packing $\phi_J=0.845$.  Because the $\alpha=0.01$ particles are to the eye indistinguishable from circles, we draw a line on each particle to indicate the direction of the particle's spine.  Animations showing the evolution of particle positions and orientations, as these systems are sheared starting from a random initial configuration,  may be found in our Supplemental Material \cite{SM}.  

While the structure and flow pattern of the particles in these animations look complex, especially for $\alpha=4$, the orientational ordering of the particles can be represented more simply by constructing a local nematic order parameter $\mathbf{S}_2(\mathbf{r})$.  To do this we divide our system up into a $12\times 12$ grid of square cells centered at fixed positions $\mathbf{r}$.  At any given strain $\gamma=\dot\gamma t$ we take all particles whose center of mass $\mathbf{r}_i$ lie in the cell at $\mathbf{r}$ and construct the local $\mathbf{S}_2$ of that cell, using Eqs.~(\ref{eS2g1}) and (\ref{eS2g2}) but with the sum  restricted to only the particles in that cell; on average there are about seven particles in each cell.   In Figs.~\ref{configs}(c) and \ref{configs}(d) we show the resulting $\mathbf{S}_2(\mathbf{r})$ corresponding to the particle configurations in \ref{configs}(a) and \ref{configs}(b).  For the $\alpha=4$ configuration, which has a relatively large global ${S}_2\approx 0.78$, we see that the $\mathbf{S}_2(\mathbf{r})$ clearly look ordered, with for the most part nearly equal magnitudes $S_2(\mathbf{r})$ and oriented close to the flow direction.  For the $\alpha=0.01$ configuration, which has a  smaller global ${S}_2\approx 0.23$, the $\mathbf{S}_2(\mathbf{r})$ look more disordered, with a greater variation in magnitudes and  directions fluctuating about the global orientation $\theta_2\approx 45^\circ$ \cite{MT2}.

Animations of the evolution of $\mathbf{S}_2(\mathbf{r})$ as $\gamma$ increases may be found in our Supplemental Material \cite{SM}.  We see in these animations that the initial $\mathbf{S}_2(\mathbf{r})$ are random, since we start in a randomized initial configuration, but that they then order as the system is sheared.  After sufficient shearing, the $\mathbf{S}_2(\mathbf{r})$ tend to fluctuate about a well defined average, and there is no evidence of any coherent time dependent motion.  Occasionally we see that $\mathbf{S}_2(\mathbf{r})$ in a given cell shrinks in size to a small value, then grows back to the average; this occurs when there is a rotation of particles in that cell.  
We now seek to quantify aspects of the spatial flow and structure by measuring the spatial correlations of several different observables.

\subsection{Flow Profile}
\label{sflow}

First we wish to check that the simple shearing in the $\mathbf{\hat x}$ direction gives rise to the  linear velocity profile, $\langle v_x(y)\rangle = \dot\gamma y$,  that is expected for a uniformly sheared system.  To compute $\langle v_x(y)\rangle$ we divide the system into  strips of thickness $\Delta y$ running the length $L_x$ of the system parallel to the flow direction.  We then compute for a given configuration,
\begin{equation}
v_x(y)=\frac{1}{N_y}\sum_{i=1}^{N_y}v_{ix},
\end{equation}
where $v_{ix}=\dot x_i$ is the $x$ component of the center of mass velocity of particle $i$, and the sum is over all the $N_y$ particles $i$ contained within the strip centered at height $y$.  On average $N_y = N\Delta y/L_y$.  We then average this over  configurations contained with in window of strain from $\gamma_0$ to $\gamma_0+\Delta\gamma$, with $\Delta\gamma=5$, to compute an average $\langle v_x(y)\rangle_{\gamma_0}$ after the system has been sheared to a strain $\gamma_0$.  We also average over all configurations in the steady-state ensemble, starting from $\gamma_0=25$ to allow for equilibration, to compute the ensemble average $\langle v_x(y)\rangle$.  We  consider here configurations sheared at a rate $\dot\gamma=10^{-6}$.

\begin{figure}
\centering
\includegraphics[width=3.5in]{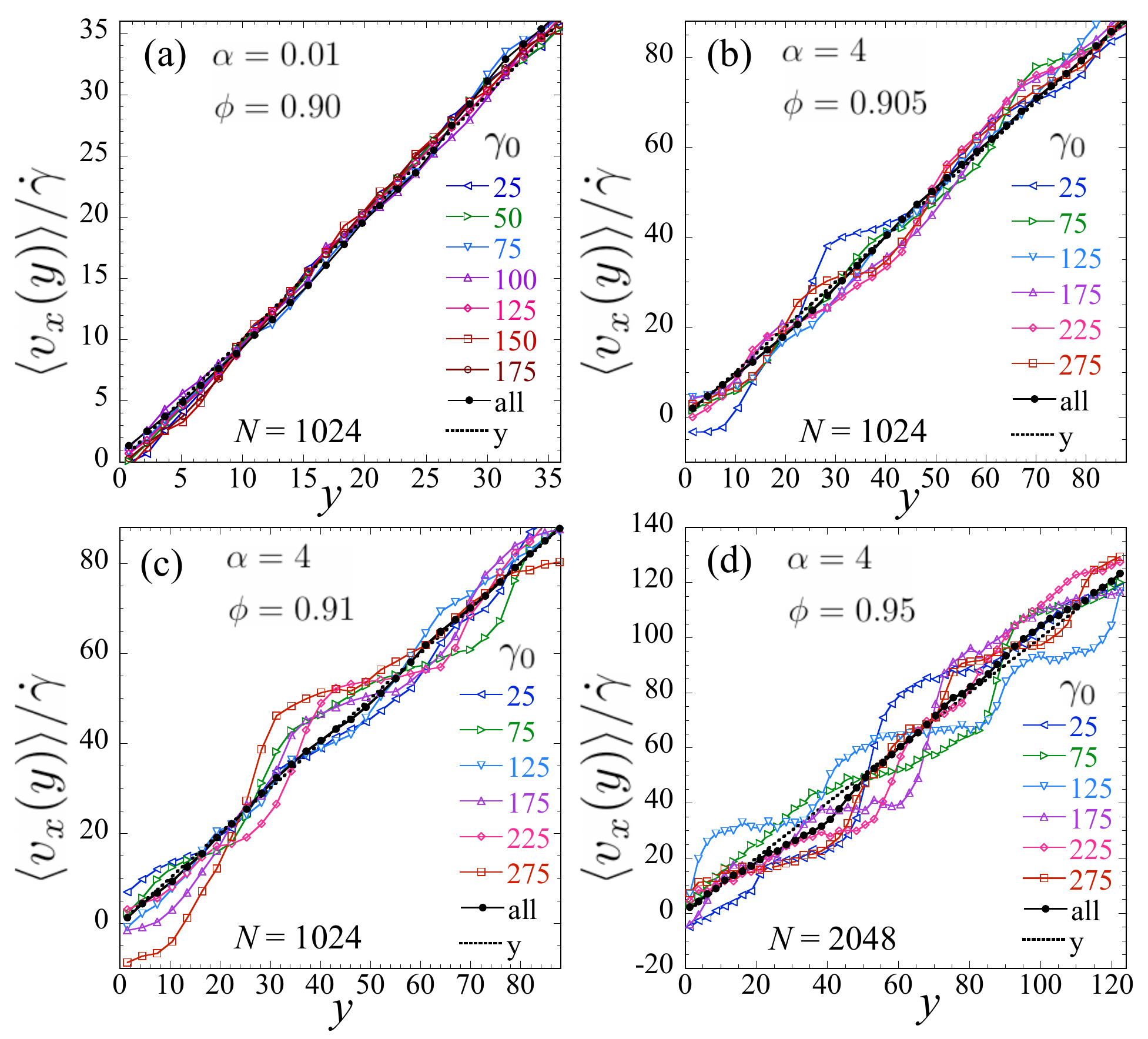}
\caption{Average velocity of particles in the flow direction scaled by the strain rate, $\langle v_x(y)\rangle/\dot\gamma$, as a function of height $y$ transverse to the flow.  Curves labeled by a value of $\gamma_0$ represent averages over a strain window from $\gamma_0$ to $\gamma_0+\Delta\gamma$, with   $\Delta\gamma=5$.  Solid black circles labeled ``all" are an average over the entire shearing run, starting at an initial $\gamma_0=25$ to allow for equilibration.  The dotted black line gives the expected linear profile $\langle v_x(y)\rangle/\dot\gamma=y$. (a) Spherocylinders with $\alpha=0.01$ at our densest packing $\phi=0.90$; (b), (c), (d) spherocylinders with $\alpha=4$ at packings $\phi=0.905\approx \phi_J$, 0.91, and 0.95 respectively.  All configurations are sheared at the rate $\dot\gamma=10^{-6}$.  Configurations (a), (b), and (c) have $N=1024$ particles, while (d) has $N=2048$ particles.  In all cases, the horizontal axis runs from 0 to $L_y$.  Lengths are measured in units of the small particle diameter, $2R_s=1$.
}
\label{vx-vs-y} 
\end{figure}

In Fig.~\ref{vx-vs-y}(a) we show our results for nearly circular spherocylinders with $\alpha=0.01$, at our densest packing $\phi=0.90$, well above the jamming $\phi_J=0.845$.  We see that the velocity profile agrees quite well with the expected linear $\langle v_x(y)\rangle/\dot\gamma=y$, both for the ensemble average over the entire run, as well as the averages over the strain windows of width $\Delta\gamma$ distributed throughout the shearing.  The same is true for all  packings at smaller $\phi$.

In Figs.~\ref{vx-vs-y}(b), \ref{vx-vs-y}(c), and \ref{vx-vs-y}(d) we show results for elongated spherocylinders with $\alpha=4$, at packings $\phi=0.905\approx \phi_J$, 0.91, and 0.95 respectively.  Note, all systems have $N=1024$ particles except for Fig.~\ref{vx-vs-y}(d) which has $N=2048$ particles.  For $\phi<\phi_J$ (not shown) the velocity profiles on the short strain scale of $\Delta\gamma=5$ are all linear, similar to what is  seen in Fig.~\ref{vx-vs-y}(a) for $\alpha=0.01$.  However, as $\phi$ increases above $\phi_J$, we see in Figs.~\ref{vx-vs-y}(b), \ref{vx-vs-y}(c), and \ref{vx-vs-y}(d), that the velocity profiles averaged over $\Delta\gamma=5$ start to noticeably fluctuate away from linear, and this effect grows in magnitude as $\phi$ increases.  We see a step-like structure, with distinct regions of different $d\langle v_x\rangle /dy$, i.e., regions of different local strain rate.  The system thus displays shear banding.  In some cases there are regions where $d\langle v_x\rangle/dy\approx 0$, indicating strongly correlated rows of particles that move together as a block, with an interface region of large strain rate between such blocks, suggesting a stick-slip type of motion between rows of particles.
However, comparing velocity profiles at different strains $\gamma_0$ during the shearing run, we see that these shear bands are not stationary, but wander as the system is sheared.  Averaging over the entire shearing run, the expected linear profile for $\langle v_x(y)\rangle$ is recovered, and so on average the system is uniformly sheared as expected.

\subsection{Transverse Velocity Correlations}

Next we consider the correlations of the transverse velocity, $v_{iy}=\dot y_i$.  It was previously found for our model  \cite{OlssonTeitelPRL}, that when circular disks are sheared, then the transverse velocity correlation 
\begin{equation}
C_{v_y}(\mathbf{r})\equiv\langle v_y(x)v_y(0)\rangle
\end{equation}
goes negative and has a minimum at some $x_\mathrm{min}$, before decaying to zero at large $x$.  It was observed that the location of this minimum $x_\mathrm{min}$ increased in a seemingly divergent way as  jamming  was approached.  Thus $x_\mathrm{min}$ was identified with the divergent correlation length $\xi$ at the jamming transition \cite{OlssonTeitelPRL}.  We now examine this velocity correlation for spherocylinders.

If $\mathbf{r}_i^c$ is the center of mass position of particle $i$ in configuration $c$, and $\mathbf{v}_i^c=\dot {\mathbf{r}}_i^c$ is the center of mass velocity, we compute the velocity correlation as,
\begin{equation}
\langle v_y(\mathbf{r})v_y(0)\rangle=\frac{1}{N_\mathbf{r}}\sum_c\sum_{i,j}v_{iy}^cv_{jy}^c\Delta(\mathbf{r}^c_i-\mathbf{r}^c_j+\mathbf{r}).
\label{eCvy}
\end{equation}
Here the first sum is over configurations $c$ in the sheared steady-state, while the second sum is over all pairs of particles $(i,j)$ in configuration $c$.  To coarse grain the point center of masses, we take $\Delta(\mathbf{r})$ as a window function, such that $\Delta(\mathbf{r})=1$  within a small square area of width $\Delta x=\Delta y = R_s =0.5$  centered about $\mathbf{r}=0$,  and $\Delta(\mathbf{r})=0$ elsewhere.  $N_\mathbf{r}$ is  the total number of non-zero terms in the sum.

\begin{figure}
\centering
\includegraphics[width=3.5in]{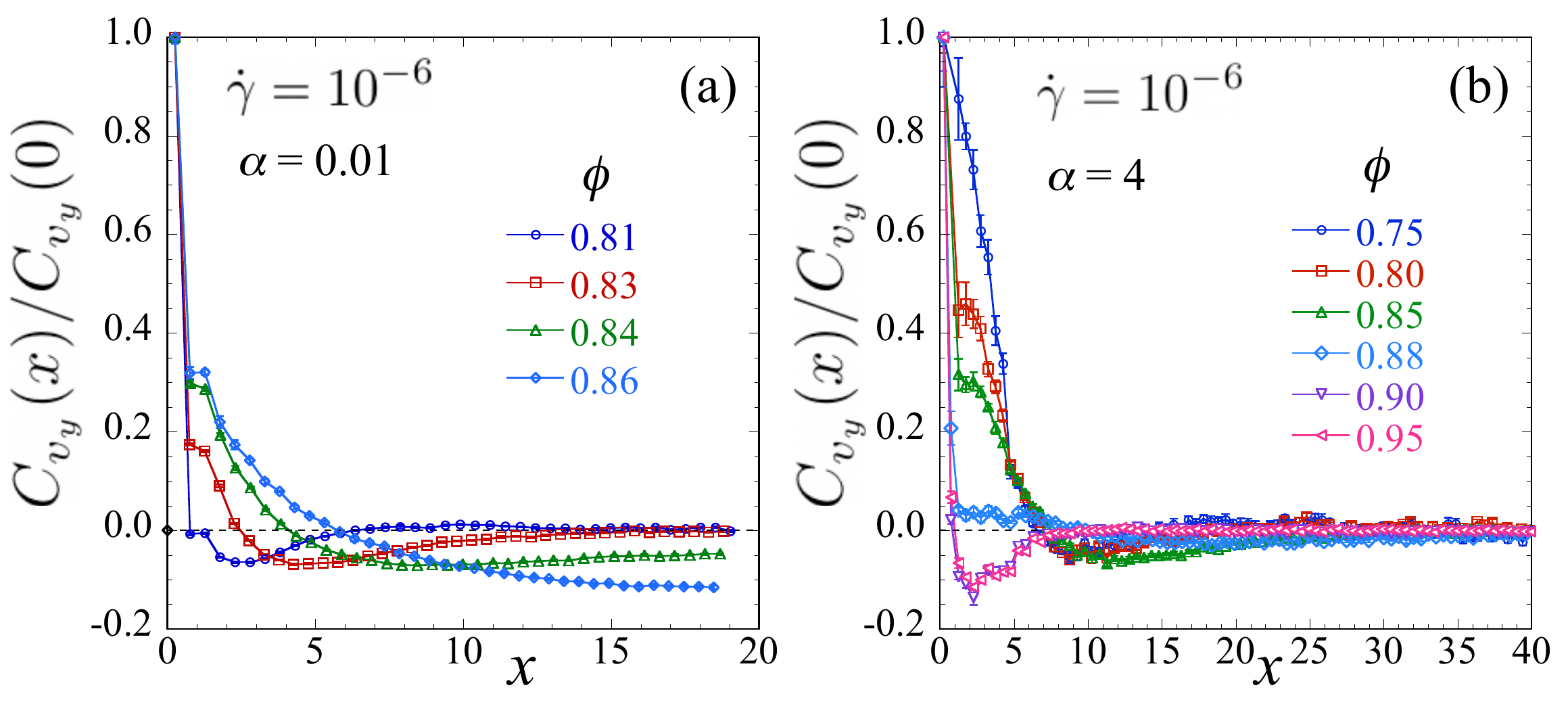}
\caption{Transverse velocity correlation $C_{v_y}(x)/C_{v_y}(0)$ vs displacement $x$ parallel to the shear flow, at different  packing fractions $\phi$ for spherocylinders of asphericity (a) $\alpha=0.01$ with $\phi_J=0.845$ and system length $L\approx 40$, and (b) $\alpha=4$  with $\phi_J=0.906$ and system length $L\approx 90$.  Both systems are sheared at a strain rate $\dot\gamma=10^{-6}$ and have $N=1024$ particles.  Lengths are measured in units of the small particle diameter $2R_s=1$.
}
\label{Cvy} 
\end{figure}

Setting $\mathbf{r}=x\mathbf{\hat x}$, we show our results in Fig.~\ref{Cvy} for nearly circular spherocylinders with $\alpha=0.01$ and moderately elongated spherocylinders with $\alpha=4$, considering different packing fractions $\phi$, below, near to, and above $\phi_J$; our results are for a strain rate $\dot\gamma=10^{-6}$.  In order to more easily compare correlations at different packings $\phi$, we show the normalized correlation $C_{v_y}(x)/C_{v_y}(0)$ vs $x$. For $\alpha=0.01$, shown in Fig.~\ref{Cvy}(a), we see behavior similar to that found \cite{OlssonTeitelPRL} for circular particles.  The correlation shows a clear minimum at an $x_\mathrm{min}$ that increases as $\phi$ approaches $\phi_J$.  Above $\phi_J$ this $x_\mathrm{min}$ increases to $L_x/2$, indicating long range transverse velocity correlations.

For the elongated particles with $\alpha=4$, shown in Fig.~\ref{Cvy}(b), the situation is quite different.  At small $\phi$, behavior is similar to $\alpha=0.01$, with a minimum at an $x_\mathrm{min}$ that increases as $\phi$ increases.  However, as the packing increases above $\phi\approx 0.88$, but still below the jamming $\phi_J=0.906$, the behavior changes dramatically with $x_\mathrm{min}$ suddenly decreasing from $x_\mathrm{min}\approx 18$ to $x_\mathrm{min}\approx 2$, and the correlations staying quite flat and zero for $x\gtrsim 10$.  Increasing $\phi$ further, to jamming and above, results in little further change in $C_{v_y}(x)/C_{v_y}(0)$.  

\begin{figure}
\centering
\includegraphics[height=1.5in]{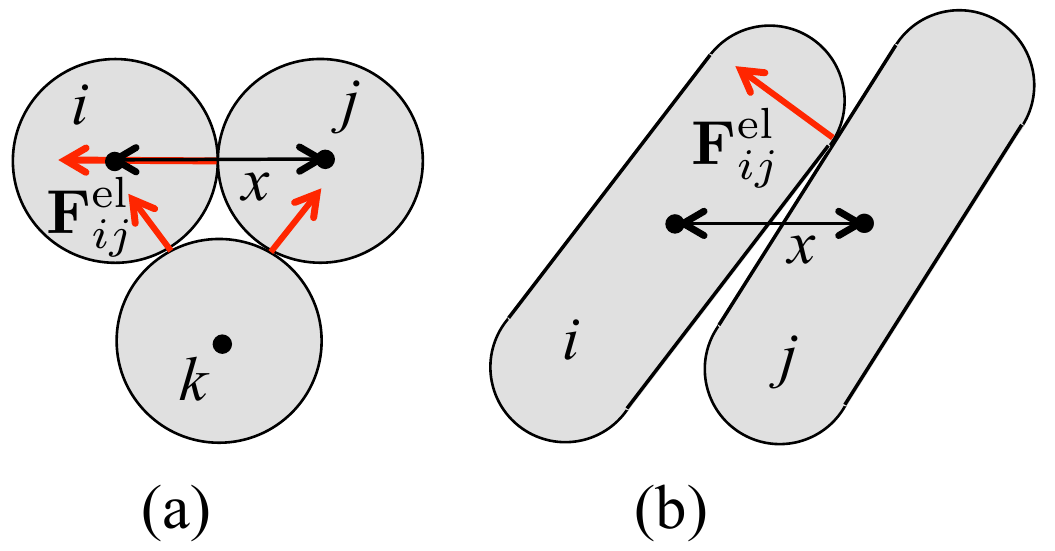}
\caption{(a) Sketch of a configuration of nearly circular spherocylinders that contributes to $C_{v_y}(x)/C_{v_y}(0)$ for small $x$.  The contact force between particles $i$ and $j$ is in the $\mathbf{\hat x}$ direction; any correlation in the $v_y$ components of the velocities of $i$ and $j$ must therefore come from contact with a third particle $k$, and gives a positive correlation.  (b) Sketch of a configuration of elongated spherocylinders that contributes to $C_{v_y}(x)/C_{v_y}(0)$ for small $x$.  Now the contact force between $i$ and $j$ will have a component in the $\mathbf{\hat y}$ direction, and so lead to a negative correlation between the $v_y$ of $i$ and $j$ since $\mathbf{F}_{ij}^\mathrm{el}=-\mathbf{F}_{ji}^\mathrm{el}$.
}
\label{Cvypic} 
\end{figure}

The difference in behavior at small $x\lesssim 2R_s$, between $\alpha=0.01$ and 4, can partially be understood as an effect of the change in particle shape. 
For small $x$, of order the particle size, $C_{v_y}(x)/C_{v_y}(0)$ is determined by contacts between particles whose centers of mass are separated by $x\mathbf{\hat x}$.  Since the force is always directed normal to the particle's surface, for circular and nearly circular particles this force is typically closely aligned with the $\mathbf{\hat x}$ direction, and so by itself induces no correlation in the $v_y$ components of the two particles' velocities.  Any correlation in $v_y$ between these two particles presumably comes from a third particle in contact with both, either from above or below, as illustrated in Fig.~\ref{Cvypic}(a), and so leads to a positive correlation.  For two elongated spherocylinders, however, if the particles are oriented at some finite angle $\theta_i>0$, then the force of the two contacting particles has a finite component in the $\mathbf{\hat y}$ direction, leading to an anti-correlation in the $v_y$ components of the two particles' velocities, as illustrated in Fig.~\ref{Cvypic}(b).  This explains the negative values of $C_{v_y}(x)/C_{v_y}(0)$ at small $x$, seen in Fig.~\ref{Cvy}(b).  However, we have no clear understanding why this effect for $\alpha=4$ seems to only occur for $\phi > 0.88$, or why for $\phi>0.88$ the correlation $C_{v_y}(x)/C_{v_y}(0)$ becomes quite flat, and shows no other structure for $x\gtrsim 5$.

We note that the identification of $x_\mathrm{min}$ with a diverging correlation length $\xi$ has recently been questioned \cite{OTunp}.  Were $x_\mathrm{min}\propto\xi$, one would expect that a scaled $C_{v_y}(x)/C_{v_y}(0)$, when plotted vs $x/x_\mathrm{min}$ at different $\phi$ or $\dot\gamma$, would show a collapse to a common curve at large $x/x_\mathrm{min}$.  But, for circular particles, this has been found not to be the case; rather the minimum at $x_\mathrm{min}$ is now believed to be a consequence of competition between two different length scales.  One should therefore not take the results of Fig.~\ref{Cvy}(b) as clear evidence for the absence of a diverging $\xi$ for $\alpha=4$, and indeed the critical scaling analysis of pressure that we have recently done for $\alpha=4$ \cite{MT1} suggests that such a diverging $\xi$ does indeed exist, although it is apparently not evident in the transverse velocity correlations.

\subsection{Positional Correlations}
\label{sPC}

For  spherical particles, it is observed that there is no long range translational ordering when the particles are sheared  \cite{Sastry}.  Since our spherocylinders do show orientational ordering when sheared, it is of interest to see if such orientational ordering might induce any translational ordering. 
We therefore consider the positional correlations of the particles, to confirm that there is no such translational ordering.  With the average particle density given by $n_0\equiv N/L^2$, we define the density-density correlation function as,
\begin{equation}
C_n(\mathbf{r})=\frac{1}{n_0^2}\left[\langle n(\mathbf{r})n(0)\rangle - n_0^2\right].
\end{equation}

To evaluate $C_n(\mathbf{r})$, we compute the ensemble average,
\begin{equation}
C_n(\mathbf{r})=\frac{1}{n_0^2}\left\langle \frac{1}{L^2}\sum_{i,j}\delta (\mathbf{r}_i-\mathbf{r}_j+\mathbf{r})\right\rangle -1,
\end{equation}
where in practice the $\delta (\mathbf{r})$ is smeared out over a small bin of area $\Delta a$ centered at the origin, so that $\delta(\mathbf{r})=0$ outside the bin and $1/\Delta a$ within the bin; the width of the bin is roughly $\sqrt{\Delta a}\approx 0.1$ for $\alpha=0.01$ and $\sqrt{\Delta a}\approx 0.2$ for $\alpha=4$, where $R_s=0.5$ is the radius of the small particles.  The finite width of our bins will affect the heights and fine structure of the sharp peaks in $C_n(\mathbf{r})$ that occur at separations corresponding to neighboring particle contacts, but otherwise does not effect the large $|\mathbf{r}|$ behavior that is  our interest here.
With the normalization we have chosen, our density correlation $C_n(\mathbf{r})$  is simply related to the usual pair correlation function $g(\mathbf{r})$ by,
\begin{equation}
g(\mathbf{r})=C_n(\mathbf{r})+1.
\end{equation}

\begin{figure}
\centering
\includegraphics[width=3.5in]{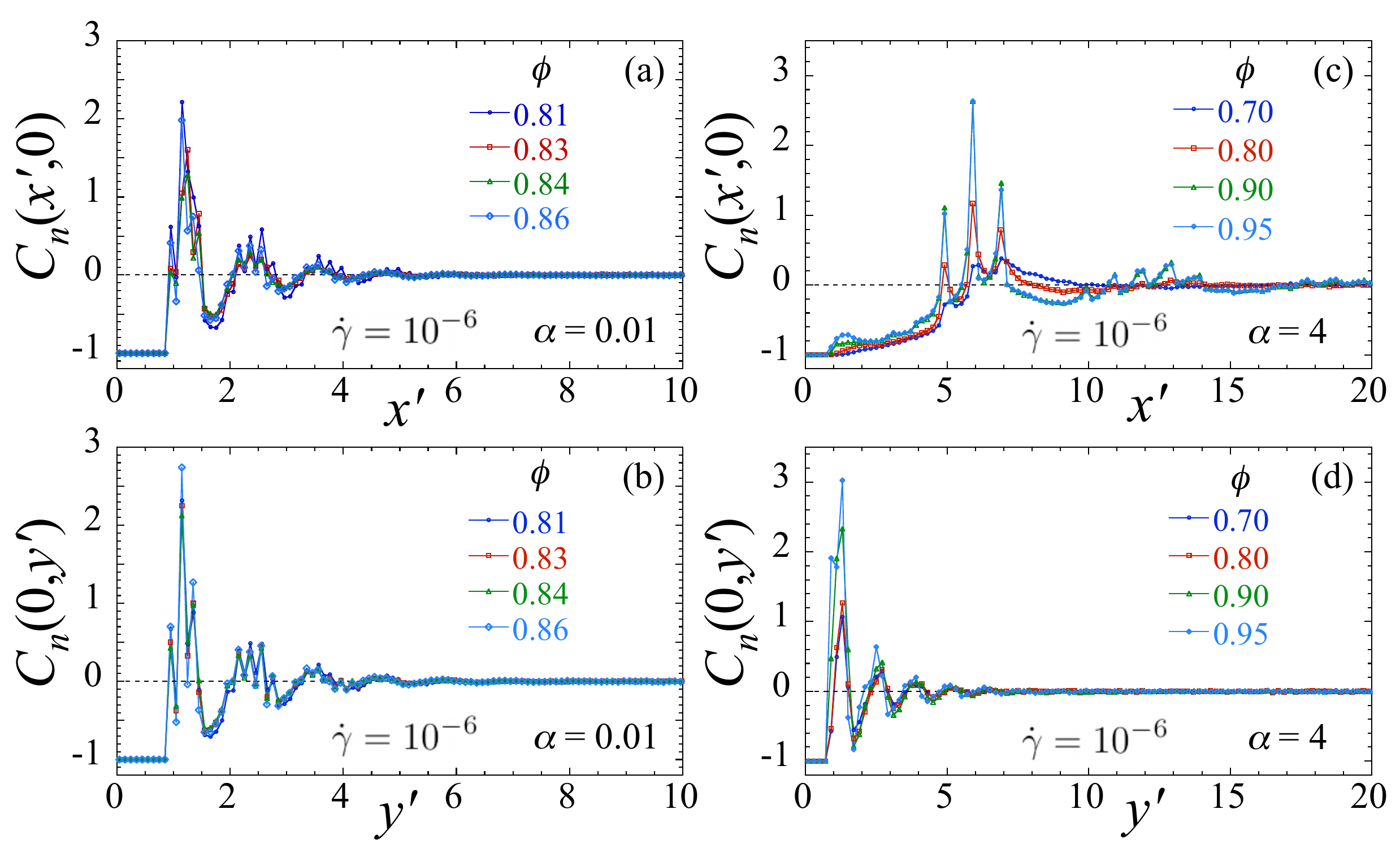}
\caption{Density correlation $C_n(\mathbf{r})$ vs coordinates $x^\prime$ and $y^\prime$, parallel and perpendicular to the nematic order parameter $\mathbf{S}_2$, at different packing fractions $\phi$.   (a) and (b) are for spherocylinders of $\alpha=0.01$, with $\phi_J=0.845$ and system length $L\approx 40$; (c) and (d) are for $\alpha=4$, with $\phi_J=0.906$ and $L\approx 90$.  Both systems are sheared at a strain rate $\dot\gamma=10^{-6}$ and have $N=1024$ particles.  Lengths are measured in units of the small particle diameter, $2R_s=1$.
}
\label{Cn} 
\end{figure}

Because the rotational symmetry of the system is broken by both the flow direction $\mathbf{\hat x}$ and by the direction of the nematic order parameter $\mathbf{S}_2$,  the correlation $C_n(\mathbf{r})$  will not be rotationally invariant.  Therefore, instead of averaging over orientations and plotting as a function of the radial coordinate, as is often done, we will instead consider separately the behavior of $C_n(\mathbf{r})$ in orthogonal directions.   One choice would be to look along the $x$ and $y$ directions, parallel and transverse to the shear flow.  However, since individual particles tend to align  parallel to $\mathbf{S}_2$,  we consider instead the direction oriented parallel to $\mathbf{S}_2$, which we denote as $x^\prime$, and the orthogonal direction, which we denote as $y^\prime$. Writing $\mathbf{r}=(x^\prime, y^\prime)$, in Fig.~\ref{Cn} we plot $C_n(x^\prime,0)$ vs $x^\prime$, and $C_n(0,y^\prime)$ vs $y^\prime$, for spherocylinders of asphericity $\alpha=0.01$ and $\alpha=4$.  We show results at several different packings $\phi$, below, near to, and above $\phi_J$; our results are for a strain rate $\dot\gamma=10^{-6}$.

For the nearly circular particles with $\alpha=0.01$, shown in Figs.~\ref{Cn}(a) and \ref{Cn}(b), we see little difference between the $x^\prime$ and $y^\prime$ directions, or among the different $\phi$.  Fitting the peak heights to an exponential decay, we find that the correlation $C_n(\mathbf{r})$  decays to zero on a length scale $\approx 1$, much shorter than the system half length, $L/2\approx 20$.  We see that $C_n(\mathbf{r})=-1$ for $x^\prime,y^\prime\lesssim 1$, since no particles may come closer to each other than $2R_s=1$ without an unreasonable particle overlap.  We see the  nearest neighbor peak is split into three at distances $x^\prime, y^\prime\approx 1.0,$ 1.2, and 1.4, corresponding to contacts between small-small, small-big, and big-big particles.

For the elongated particles with $\alpha=4$, shown in Figs.~\ref{Cn}(c) and \ref{Cn}(d), however, we see a big difference between the $x^\prime$ and $y^\prime$ directions.  Since the $(x^\prime,y^\prime)$ coordinates are aligned parallel and perpendicular to $\mathbf{S}_2$, and since particles on average are also aligned with their spines parallel to $\mathbf{S}_2$, the $x^\prime$ coordinate on average runs parallel to the particle spines. Therefore, for parallel oriented particles aligned in a row, the closest approach another particle can make in the $x^\prime$ direction is the length of a small particle, $2R_s(\alpha+1)=5$, and hence in Fig.~\ref{Cn}(c) we see the nearest neighbor peaks at $x^\prime\approx 5$, 6, and 7, corresponding to nearest contacts between small-small, small-big, and big-big particles.  In the transverse $y^\prime$ direction, however, corresponding to the narrow width of the particle, the closest parallel oriented particles aligned in a row may come is $2R_s=1$.  In principle, we would expect to see peaks at $y^\prime=1$, 1.2 and 1.4, corresponding to small-small, small-big, and big-big particle contacts, however the finite size of our bins (which are a bit larger here than for $\alpha=0.01$) make these less sharply distinguished.

Note, for $\alpha=4$, the correlation $C_n(0,y^\prime)$ drops sharply to $-1$ as $y^\prime$ decreases below unity.  This is because the shortest distance any two particles may approach each other, without unreasonable overlaps, is $2R_s=1$.  However for $C_n(x^\prime,0)$ we see no such sharp drop as $x^\prime$ decreases below $2R_s(\alpha+1)=5$.  In fact, $C_n(x^\prime,0)$ becomes, and stays equal to, $-1$ only when $x^\prime$ decreases below $2R_s=1$.  The reason for this is that not all particles are aligned nearly parallel to $\mathbf{S}_2$.  When two adjacent particles are  aligned nearly perpendicular to $\mathbf{S}_2$, then one can have a contact at $x^\prime\approx 1$; although this is possible  (see Fig.~\ref{configs}(a)), it is relatively uncommon, hence $C_n(\mathbf{r})$ increases slowly above $-1$ as $x^\prime$ increases above unity, then takes a rapid increase at $x^\prime\approx 5$.  This lack of perfect alignment of particles parallel to $\mathbf{S}_2$ is also responsible for the the fact that the sharp peaks in Fig.~\ref{Cn}(c) are not exactly at $x^\prime=5$, 6, and 7, but rather are  at slightly smaller values.

Comparing the $\phi$ dependence of $C_n(\mathbf{r})$ for $\alpha=4$, we see little effect in the transverse direction $y^\prime$, but in the $x^\prime$ direction one sees more clearly higher order peaks as $\phi$ approaches and goes above  $\phi_J$.  In all cases, however, $C_n(\mathbf{r})$ decays to zero as $|\mathbf{r}|$ increases; for the $y^\prime$ direction the decay length is $\approx 1.3$, while in the $x^\prime$ direction it is $\approx 4$.

\begin{figure}
\centering
\includegraphics[width=3.5in]{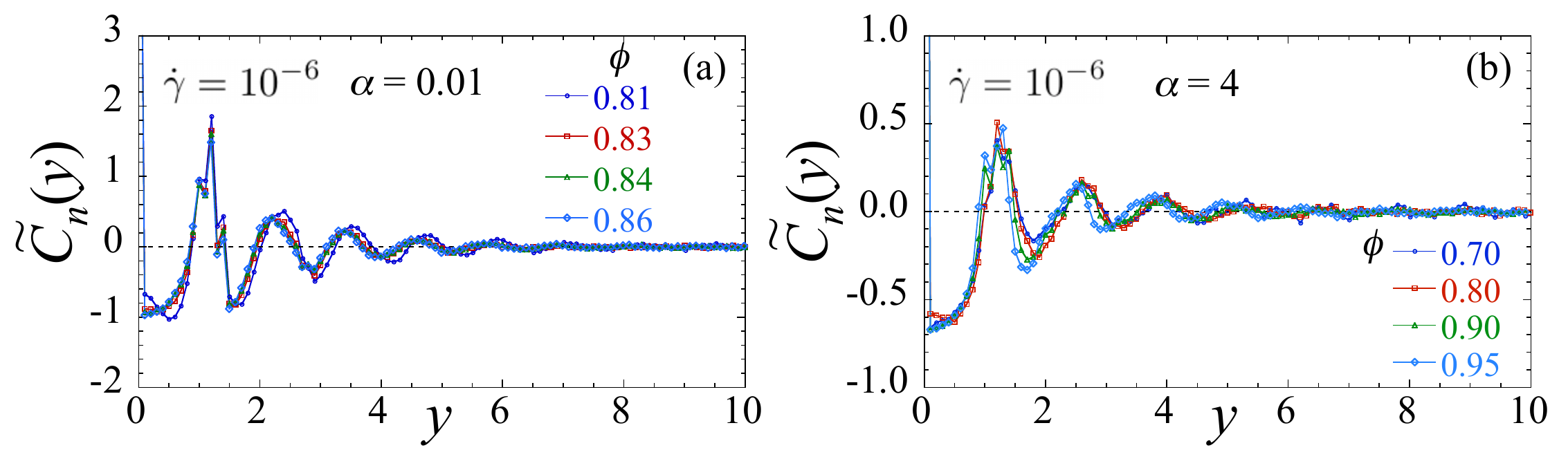}
\caption{Transverse density correlation $\tilde C_n(y)$ vs $y$, at different packing fractions $\phi$, for spherocylinders of  (a) $\alpha=0.01$, with $\phi_J=0.845$ and system length $L\approx 40$; (b) $\alpha=4$, with $\phi_J=0.906$ and $L\approx 90$.  Both systems are sheared at a strain rate $\dot\gamma=10^{-6}$ and have $N=1024$ particles.  Lengths are measured in units of the small particle diameter, $2R_s=1$.
}
\label{Cn-lanes} 
\end{figure}

The above calculations show that the particles have no long range translational order in the sheared system.  However we still wish to investigate if there can be any shear induced columnar-like ordering, where particles order into well defined channels oriented parallel to the flow direction $\mathbf{\hat x}$.  To investigate this we average the $C_n(\mathbf{r})$ correlation over the $x$ direction to define the transverse density correlation function $\tilde C_n(y)$,

\begin{equation}
\tilde C_n(y) =n_0 \int_0^L\!\!dx\, C_n(x,y).
\end{equation}
Our results are shown in Fig.~\ref{Cn-lanes} for spherocylinders of $\alpha=0.01$ and $4$.  Again we see that these correlations rapidly decay to zero as the separation $y$ increases.  
Fitting the peak heights to an exponential gives a decay length between 1 and 2.  Thus we conclude that the particles do not flow in well defined channels and there is no columnar ordering.

\subsection{Nematic Correlations}
\label{sNC}

Next we wish to consider the correlations of the nematic order parameter $\mathbf{S}_2$.  Shearing induces a finite $\mathbf{S}_2$ in the system at any $\phi$, as shown in Fig.~\ref{S2-vs-phi}, but our arguments in Ref.~\cite{MT2}  suggested that this finite $\mathbf{S}_2$ arises because the shearing acts like an ordering field, rather than because of many-particle cooperative behavior arising from a long range coherence of particle orientations.  Computing the correlations of the nematic order parameter $\mathbf{S}_2$  will confirm this.

The nematic correlation function is,
\begin{equation}
C_{S_2}(\mathbf{r})=\langle \cos 2[\theta(\mathbf{r})-\theta(0)]\rangle - S_2^2,
\end{equation}
where the first term is computed similarly to Eq.~(\ref{eCvy}).  If $\theta_i^c$ is the orientation of particle $i$ in configuration $c$, then
\begin{equation}
\langle \cos 2[\theta(\mathbf{r})-\theta(0)]\rangle = \frac{1}{N_\mathbf{r}}\sum_c\sum_{i,j}\cos 2(\theta_i^c-\theta_j^c)\Delta(\mathbf{r}_i^c-\mathbf{r}_j^c+\mathbf{r}),
\end{equation}
where $\Delta(\mathbf{r})$ is the same window function as used in computing $C_{v_y}(x)$, and $N_\mathbf{r}$ is the number of non-zero terms being summed.

\begin{figure}
\centering
\includegraphics[width=3.5in]{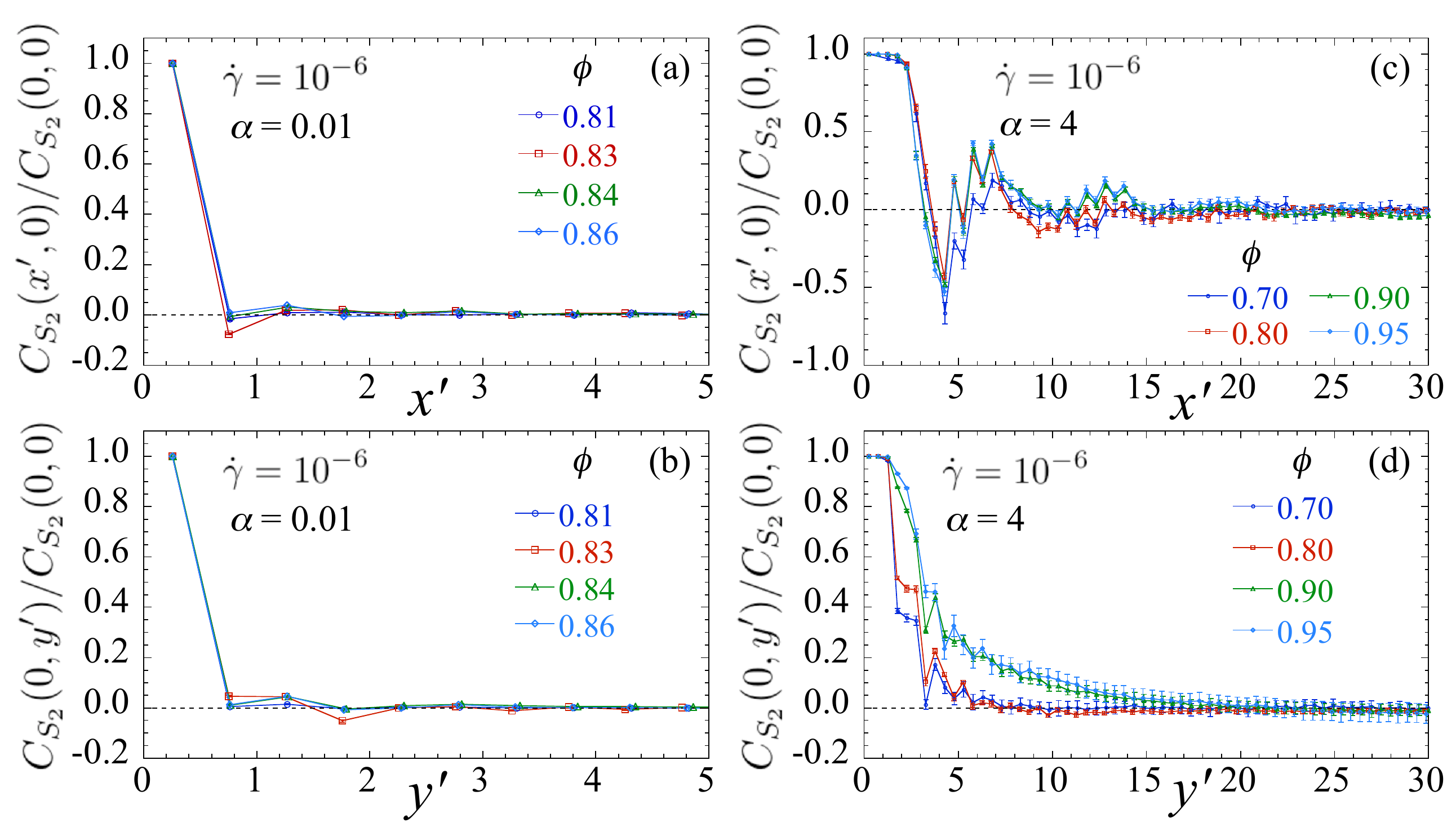}
\caption{Nematic ordering correlation $C_{S_2}(\mathbf{r})/C_{S_2}(0)$ vs coordinates $x^\prime$ and $y^\prime$, parallel and perpendicular to the global nematic order parameter $\mathbf{S}_2$, at different packing fractions $\phi$.  (a) and (b) are for spherocylinders of $\alpha=0.01$, with $\phi_J=0.845$ and system length $L\approx 40$; (c) and (d) are for $\alpha=4$, with $\phi_J=0.906$ and $L\approx 90$.  Both systems are sheared at a strain rate $\dot\gamma=10^{-6}$ and have $N=1024$ particles.  Lengths are measured in units of the small particle diameter, $2R_s=1$.
}
\label{CS2} 
\end{figure}

In Fig.~\ref{CS2} we show our results for $C_{S_2}(\mathbf{r})/C_{S_2}(0)$ in the  $x^\prime$ and $y^\prime$ directions, parallel and perpendicular to the global nematic order parameter $\mathbf{S}_2$.  We show results for different packings $\phi$, below, near to, and above $\phi_J$, for systems sheared with strain rate $\dot\gamma=10^{-6}$.
For nearly circular particles with $\alpha=0.01$, shown in Figs.~\ref{CS2}(a) and \ref{CS2}(b), we see that there is little difference in the correlation function comparing the different packings $\phi$, or comparing the $x^\prime$ and $y^\prime$ directions,  and that  the correlations decay  rapidly to zero within one small particle width, $2R_s=1$.  [Note, although no two particles may come much closer than $2R_s=1$ without an unreasonable overlap, here we see a large drop at $x^\prime= y^\prime= 0.75$; this is an artifact of the finite width $\Delta x=\Delta y=0.5$ of our window  function $\Delta(\mathbf{r})$].  The very rapid decay of the correlation function, and the absence of any noticeable variation of the decay length with the packing $\phi$, indicate that there are no long-range orientational correlations between the particles.

For moderately elongated particles with $\alpha=4$, shown in Figs.~\ref{CS2}(c) and \ref{CS2}(d), we see a noticeable difference between the $x^\prime$ and $y^\prime$ directions.  Along the $x^\prime$ direction $C_{S_2}(\mathbf{r})/C_{S_2}(0)$ is a decaying oscillation  with a period of roughly $\approx 6$, corresponding to the average length of the particles.  A rough estimate gives a decay length of comparable size $\approx 5$.  Along the $y^\prime$ direction correlations remain positive, and we see that the decay length takes a noticeable increase as $\phi$ increases,  from roughly $\approx 1.5$ at $\phi=0.80$ to $\approx 5$ at $\phi=0.90$ and above.  
Indeed for the packing $\phi=0.905$, shown in Fig.~\ref{configs}(a) and the corresponding animation \cite{SM}, it is easy to see that one has many local stacks of particles in side-to-side contact along their flat edges, nearly one on top of the other.  Such local stackings presumably result from the system adjusting to reduce the pressure at a given packing.
These stacks, often consisting of  $\sim 10$ or more particles, are then responsible for the larger decay length in the $y^\prime$ direction as $\phi$ increases above jamming.   Nevertheless, despite this increase in decay length as $\phi$ increases, the decay length appears to remain finite at all $\phi$,  $C_{S_2}(\mathbf{r})/C_{S_2}(0)$ decays to zero on the order of a typical particle size as $|\mathbf{r}|$ increases, and we thus conclude that there are no long-range orientational correlations between the particles.

\subsection{Angular Velocity Correlations}

Finally we consider the correlations of the scaled angular velocity, $\theta^\prime_i=d\theta_i/d\gamma=\dot\theta_i/\dot\gamma$,
\begin{equation}
C_{\theta^\prime}(\mathbf{r})=\left[\langle\dot\theta(\mathbf{r})\dot\theta(0)\rangle - \langle \dot\theta_i\rangle^2\right]/\dot\gamma^2.
\end{equation}
As we have done for other quantities, if $\dot\theta_i^c$ is the angular velocity of particle $i$ in configuration $c$, then we compute
\begin{equation}
\langle\dot\theta(\mathbf{r})\theta(0)\rangle = \frac{1}{N_\mathbf{r}}\sum_c\sum_{i,j}\dot\theta_i^c\dot\theta_j^c\Delta(\mathbf{r}_i^c-\mathbf{r}_j^c+\mathbf{r}).
\end{equation}

\begin{figure}
\centering
\includegraphics[width=3.5in]{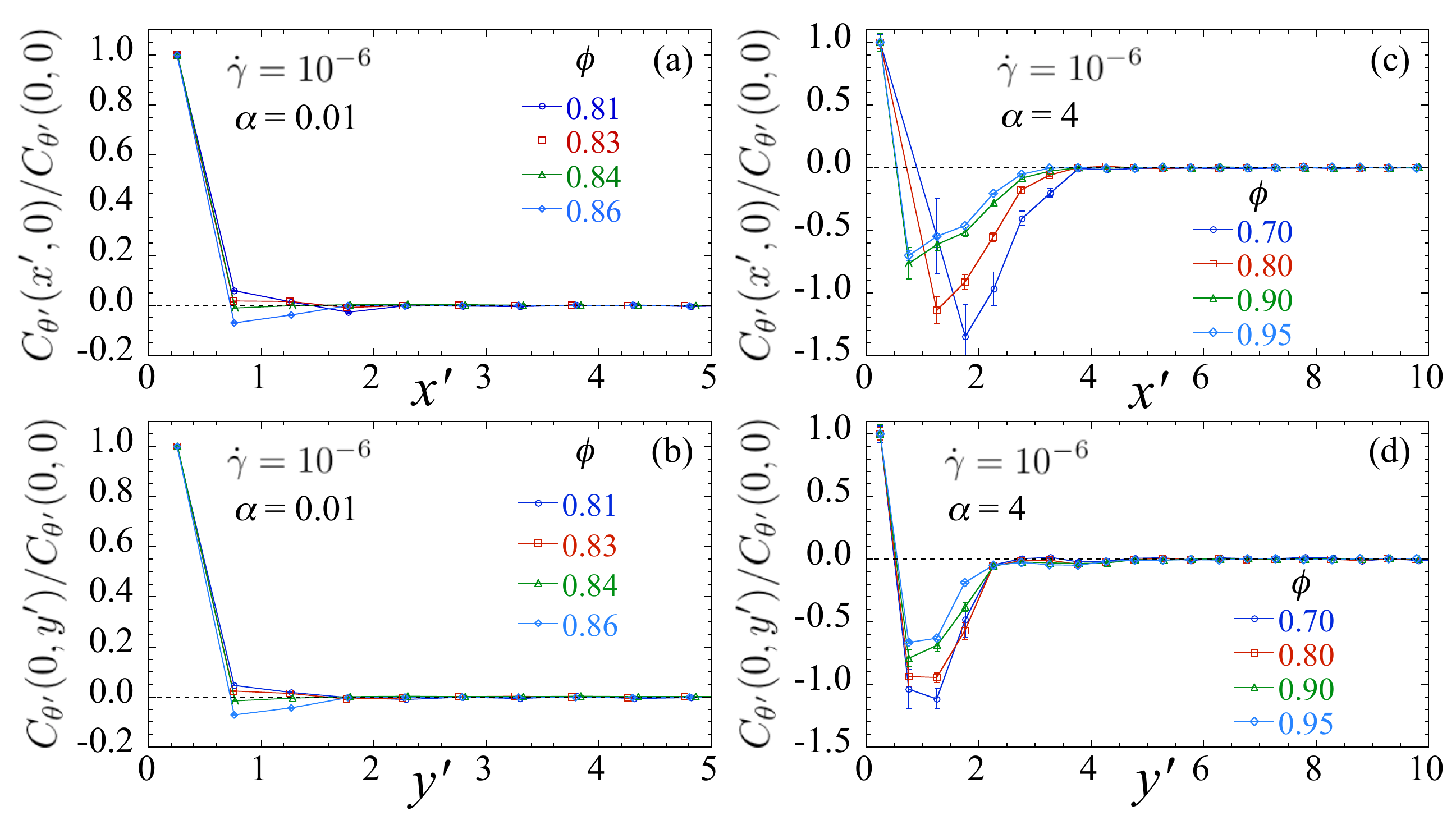}
\caption{Angular velocity correlation $C_{\theta^\prime}(\mathbf{r})/C_{\theta^\prime}(0)$, where $\theta^\prime_i=\dot\theta_i/\dot\gamma$, vs coordinates $x^\prime$ and $y^\prime$, parallel and perpendicular to the global nematic order parameter $\mathbf{S}_2$, at different packing fractions $\phi$.  (a) and (b) are for spherocylinders of $\alpha=0.01$, with $\phi_J=0.845$ and system length $L\approx 40$; (c) and (d) are for $\alpha=4$, with $\phi_J=0.906$ and $L\approx 90$.  Both systems are sheared at a strain rate $\dot\gamma=10^{-6}$ and have $N=1024$ particles.  Lengths are measured in units of the small particle diameter, $2R_s=1$. 
}
\label{Cav} 
\end{figure}

In Fig.~\ref{Cav} we show our results for $C_{\theta^\prime}(\mathbf{r})/C_{\theta^\prime}(0)$ in the  $x^\prime$ and $y^\prime$ directions, parallel and perpendicular to the global nematic order parameter $\mathbf{S}_2$.  We show results for different packings $\phi$, below, near to, and above $\phi_J$, for systems sheared with strain rate $\dot\gamma=10^{-6}$.  For both nearly circular particles with $\alpha=0.01$, shown in Figs.~\ref{Cav}(a) and \ref{Cav}(b), and for moderately elongated particles with $\alpha=4$, shown in Figs.~\ref{Cav}(c) and \ref{Cav}(d), we see that the correlation drops rapidly and stays flat at zero, once $|\mathbf{r}|$ is greater than the particle length $1+\alpha$.  Only nearest neighbor particles are at all correlated, and those are anti-correlated, as indicated by the negative value of $C_{\theta^\prime}(\mathbf{r})/C_{\theta^\prime}(0)$ at $|\mathbf{r}|\approx 1$.

\begin{figure}
\centering
\includegraphics[height=1in]{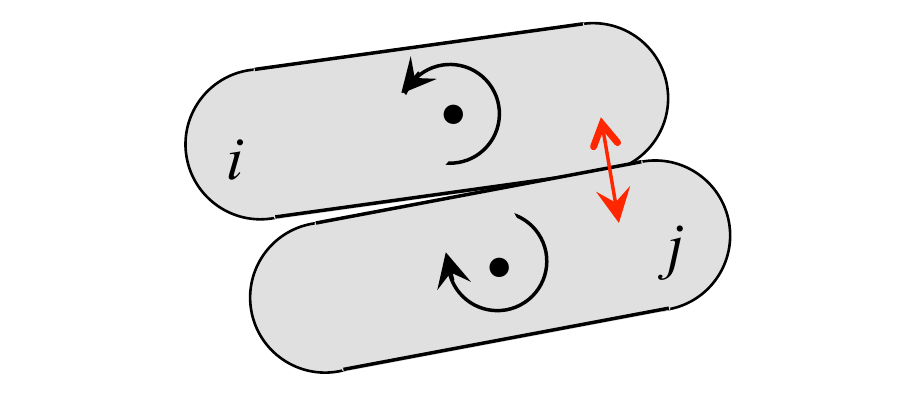}
\caption{Sketch of two nearly parallel particles to illustrate how a collision leads to oppositely oriented changes in angular velocity, and thus explains the anti-correlation seen in $C_{\theta^\prime}(\mathbf{r})$ for $|\mathbf{r}|\approx 1$.
}
\label{avspheros} 
\end{figure}

To illustrate the origin of this anti-correlation of nearest neighbor angular velocities, in Fig.~\ref{avspheros} we sketch two nearest neighbor, nearly parallel, particles with separation $|\mathbf{r}|\approx 1$. We see that a collision between the two particles, indicated by the double headed arrow in the sketch, leads to oppositely oriented changes in angular velocity for the two particles, and hence the anti-correlation.  However, for larger $|\mathbf{r}|$, on the order of a few or more particle separations, our results in Fig.~\ref{Cav} indicate that fluctuations in the particles' angular velocities are completely uncorrelated.

%%%%%%%%%%%%%%%%%%%%%%

\section{Size-Monodisperse Particles}
\label{sec:Monodisperse}

When studying jamming in two-dimensional systems of circular particles, it is common to consider bidisperse or polydisperse distributions of particle sizes, so as to avoid crystallization into an ordered hexagonal lattice.  When studying aspherically shaped particles, one can ask if the possibility of such crystallization still remains for size monodisperse particles.  In particular, for  particles driven by simple shear, the shear-driven rotation of particles could conceivably disrupt crystalline structure in densely packed systems, if the particles are sufficiently aspherical.  

\begin{figure}
\centering
\includegraphics[width=3.5in]{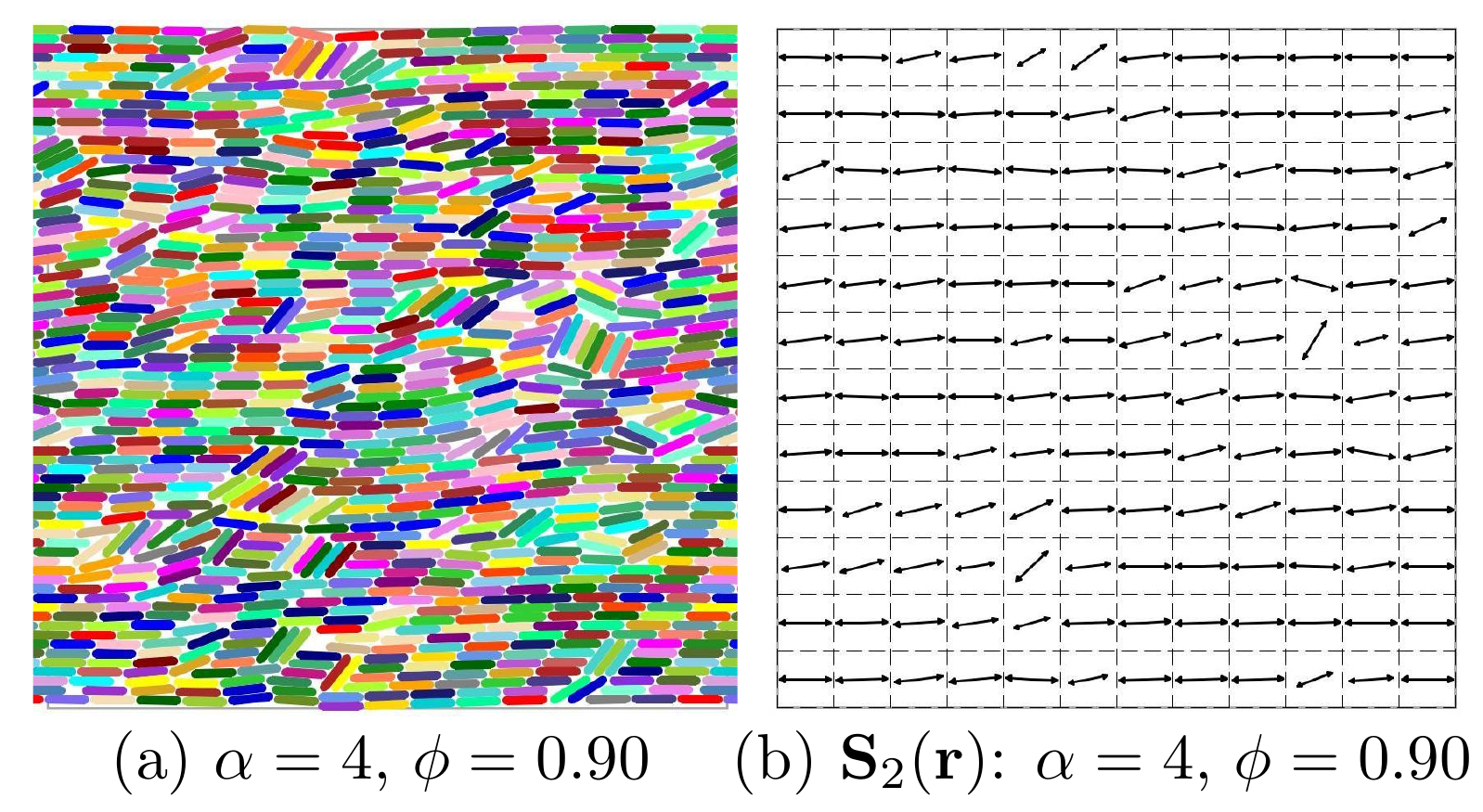}
\caption{(a) Snapshot configuration of a system of size-monodisperse particles of asphericity $\alpha=4$, at packing $\phi=0.90$, sheared at $\dot\gamma=10^{-6}$.  Different colors are used to help distinguish different particles and have no other meaning. (b) The corresponding configuration of the local nematic order parameter $\mathbf{S}_2(\mathbf{r})$, obtained by averaging over all particles whose center of mass $\mathbf{r}_i$ is contained in each square grid cell.  A corresponding animation, showing the evolution of this configuration as it is sheared, is available in our Supplemental Material \cite{SM}.
}
\label{mono-config} 
\end{figure}

In this section, therefore, we study the case of a size monodisperse system of moderately elongated spherocylinders with asphericity $\alpha=4$.
For the bidisperse distribution of $\alpha=4$ spherocylinders we have previously determined \cite{MT1} the shear-driven jamming transition to be at $\phi\approx0.906$.  For the monodisperse distribution we have not carried out a similar detailed analysis to try and locate $\phi_J$ accurately.  However, by comparing the dependence of the pressure on $\phi$ and $\dot\gamma$, our crude estimate for the jamming of the monodisperse system is $\phi_J\approx 0.92$.  
In Fig.~\ref{mono-config}(a) we show a snapshot of a typical configuration sampled during steady-state shearing at packing $\phi=0.90$ and strain rate $\dot\gamma=10^{-6}$.  In Fig.~\ref{mono-config}(b) we show the corresponding configuration of the local nematic order parameter $\mathbf{S}_2(\mathbf{r})$, computed as described earlier in connection with Fig.~\ref{configs}.
An eyeball comparison of Figs.~\ref{mono-config}(a) and (b) with the bidisperse case in Figs.~\ref{configs}(a) and (c) suggests that  for dense monodisperse systems there is a larger probability for particles to be aligned parallel to the flow direction $\mathbf{\hat x}$.  
We will return to this matter below in Sec.~\ref{GNO}.

\subsection{Flow Profile}

We have found that reliable results for the monodisperse system are difficult to obtain much above the jamming $\phi_J\approx 0.92$, because at large packings the particles tend to lock into local configurations.  This is illustrated by considering the flow profile $\langle v_x(y)\rangle$, defined earlier in Sec.~\ref{sflow}.

\begin{figure}
\centering
\includegraphics[width=3.5in]{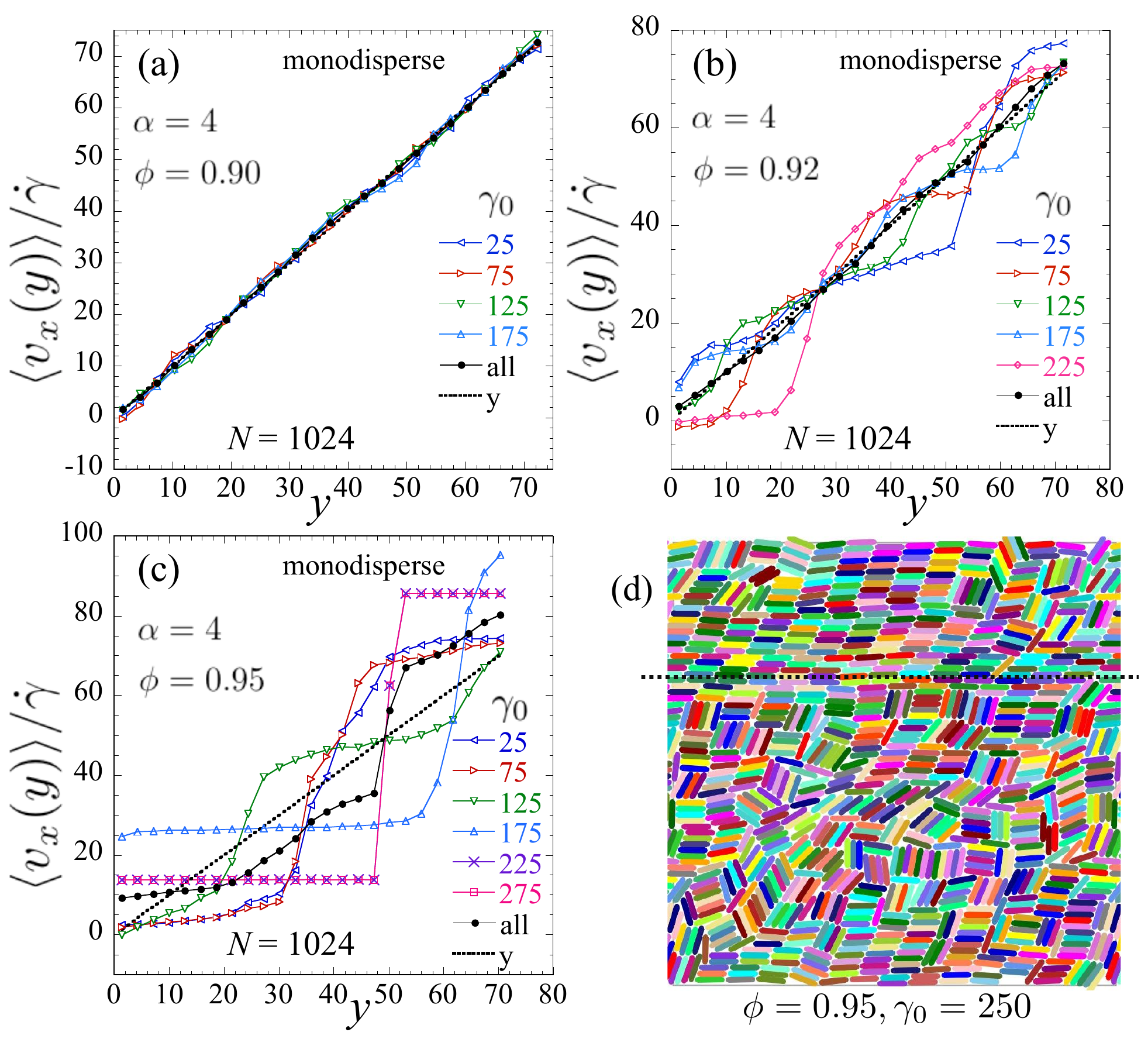}
\caption{For $N=1024$ monodisperse particles of asphericity $\alpha=4$, sheared at $\dot\gamma=10^{-6}$:  Average velocity of particles in the flow direction scaled by the strain rate, $\langle v_x(y)\rangle/\dot\gamma$, as a function of height $y$ transverse to the flow, for packing fractions (a) $\phi=0.90$, (b) $\phi=0.92$, and (c) $\phi=0.95$.  Curves labeled by a value of $\gamma_0$ represent averages over a strain window from $\gamma_0$ to $\gamma_0+\Delta\gamma$, with   $\Delta\gamma=5$.  Solid black circles labeled ``all" are an average over the entire shearing run, starting at an initial $\gamma_0=25$ to allow for equilibration.  The dotted black line gives the expected linear profile $\langle v_x(y)\rangle/\dot\gamma=y$. 
Lengths are measured in units of the small particle diameter, $2R_s=1$. In (d) is shown a snapshot of the configuration at $\phi=0.95$ after a strain of $\gamma_0=250$; the horizontal dotted line locates the interface between two coherently moving blocks of particles, as shown by the sharp jump in velocity of the corresponding curve in (c).  Different colors in (d) are used to help distinguish different particles and have no other meaning.  An animation of this configuration is available in our Supplemental Material \cite{SM}.
}
\label{flow-mono} 
\end{figure}

In Fig.~\ref{flow-mono} we show $\langle v_x(y)\rangle/\dot\gamma$ vs $y$, averaged over strain windows of width $\Delta\gamma=5$, at different total strains $\gamma_0$ within the shearing ensemble.  We also show the average over the entire shearing run.  For $\phi=0.90$, shown in Fig.~\ref{flow-mono}(a), we see that the flow profile $\langle v_x(y)/\rangle/\dot\gamma$ is almost perfectly linear for all strain windows, indicating that the shear flow is uniform even on short strain scales.  For $\phi=0.92$ near jamming, shown in Fig.~\ref{flow-mono}(b), we see the step-like structure indicative of shear banding on short strain scales; however, the location and size of these steps fluctuate with $\gamma_0$,  and when averaging over the entire run we regain the expected linear flow profile.  

However, for $\phi=0.95$, above jamming, something dramatically different occurs.  In the earlier part of the shearing run, we see wandering shear bands on short strain scales, similar to what is seen at the smaller $\phi=0.92$, only now with wider bands.  But, after shearing a large total strain, we see that the system separates  into two sharply defined bands, each with constant velocity, one small, one large, with a large velocity jump between them. The velocity profiles $\langle v_x(y)\rangle/\dot\gamma$ at $\gamma_0=225$ and $\gamma_0=275$ are identical, indicating that the system has locked into this particular state, characterized by two blocks of coherently flowing particles, each moving at different constant velocities, and sliding over each other along a sharply defined interface.  

In Fig.~\ref{flow-mono}(d) we show a snapshot of the configuration for $\phi=0.95$ at $\gamma_0=250$, after the system has locked into this state of coherently sliding blocks.  The interface between the two blocks of particles is indicated by the horizontal dotted line at height $y=50$. In either block there is neither more spatial nor orientational order than typical in a homogeneously shearing configuration, 
although there exist many local clusters of  particles contacting along their flat sides, oriented nearly in parallel; many of these clusters are oriented with the particle spines nearly parallel to the flow direction $\mathbf{\hat x}$, however, many are oriented at relatively large angles with respect to the flow.
Along the interface where the sliding takes place, one sees two rows of particles, oriented parallel to the flow, extending the length of the system; it is these rows, sliding one upon the other, that cause the large jump in velocity between the two blocks.  An animation of the shearing at $\phi=0.95$ is available in our Supplemental Material \cite{SM}; the animation starts after the system has already been sheared a considerable amount, but before it has locked into the state of coherently sliding blocks, which occurs around the midpoint of the animation.

\subsection{Positional Correlations}

We next consider the positional correlations in the monodisperse system, computing the correlation function $C_n(\mathbf{r})$, as defined earlier in Sec.~\ref{sPC}.
%Finally, to further investigate ordering in the monodisperse system, we consider the positional and orientation correlation functions, $C_n(\mathbf{r})$ and $C_{S_2}(\mathbf{r})$, defined in Secs.~\ref{sPC} and \ref{sNC}. 
Since the configuration shown in Fig.~\ref{mono-config} suggests (and as will be confirmed below in Fig.~\ref{Ptheta})
 that  many of the particles align near to the flow direction $\mathbf{\hat x}$, 
%rather than at $\theta_2$ along $\mathbf{S}_2$, 
here we will plot the correlation as a function of the $x$ and $y$ coordinates, parallel and perpendicular to the flow direction, rather than  the $x^\prime$ and $y^\prime$ coordinates (parallel and perpendicular to $\mathbf{S}_2$) used earlier for the bidisperse system in Sec.~\ref{sPC}.

\begin{figure}
\centering
\includegraphics[width=3.5in]{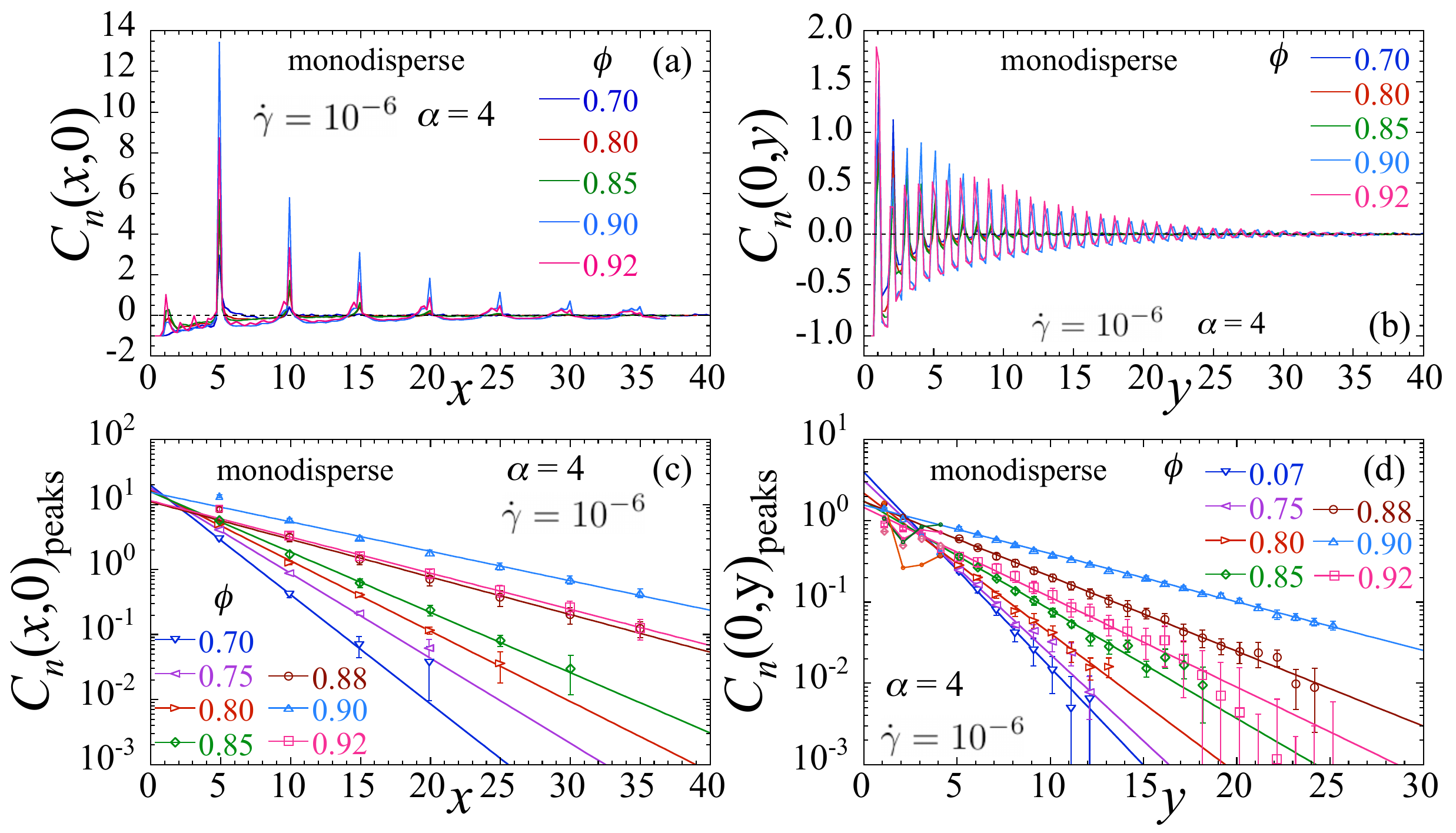}
\caption{For $N=1024$ monodisperse particles of asphericity $\alpha=4$, sheared at $\dot\gamma=10^{-6}$: Density correlation $C_n(\mathbf{r})$ vs coordinates (a) $x$ and (b) $y$, parallel and perpendicular to the flow direction $\mathbf{\hat x}$, for different packing fractions $\phi$. Peak heights in (c) $C_n(x,0)$ vs $x$ for $x_m\approx 5m$, and in (d) $C_n(0,y)$ vs $y$ for $y_m\approx m$; straight lines are fits to an exponential decay.  Lengths are measured in units of the particle diameter, $2R=1$ and the system width is $L\approx 90$.
}
\label{Cn-mono} 
\end{figure}

In Figs.~\ref{Cn-mono}(a) and \ref{Cn-mono}(b) we show $C_n(\mathbf{r})$ vs $x$ and $y$, respectively, at several different values of the packing $\phi$ for a system strained at the rate $\dot\gamma=10^{-6}$.  Comparing to Figs.~\ref{Cn}(c) and \ref{Cn}(d) for a bidisperse system, we see that in the monodisperse system  the peaks in both the $x$ and $y$  directions are more sharply defined and persist out to considerably longer length scales.  
Similar results have been suggested in simulations comparing monodisperse and polydisperse spherocylinders in three dimensions, for a model in which energy dissipation is by inelastic particle collisions rather than the viscous drag we use here \cite{Somfai}.

In Figs.~\ref{Cn-mono}(a) and \ref{Cn-mono}(b) the peaks are perfectly periodic with a spacing $\Delta x=5$ along the $x$ direction, and $\Delta y=1$ along the $y$ direction.  Nevertheless, the peak heights still decay exponentially with distance, as is seen in Figs.~\ref{Cn-mono}(c) and \ref{Cn-mono}(d) where we plot just the peak heights at $x_m\approx 5m$ and $y_m\approx m$ on a semi-log plot (we note that the locations of these peaks are not exactly at integer values of $x$, but are very close to them).  The straight lines in these figures are fits to an exponential decay, and we see reasonably good agreement.

\subsection{Nematic Correlations}

\begin{figure}
\centering
\includegraphics[width=3.5in]{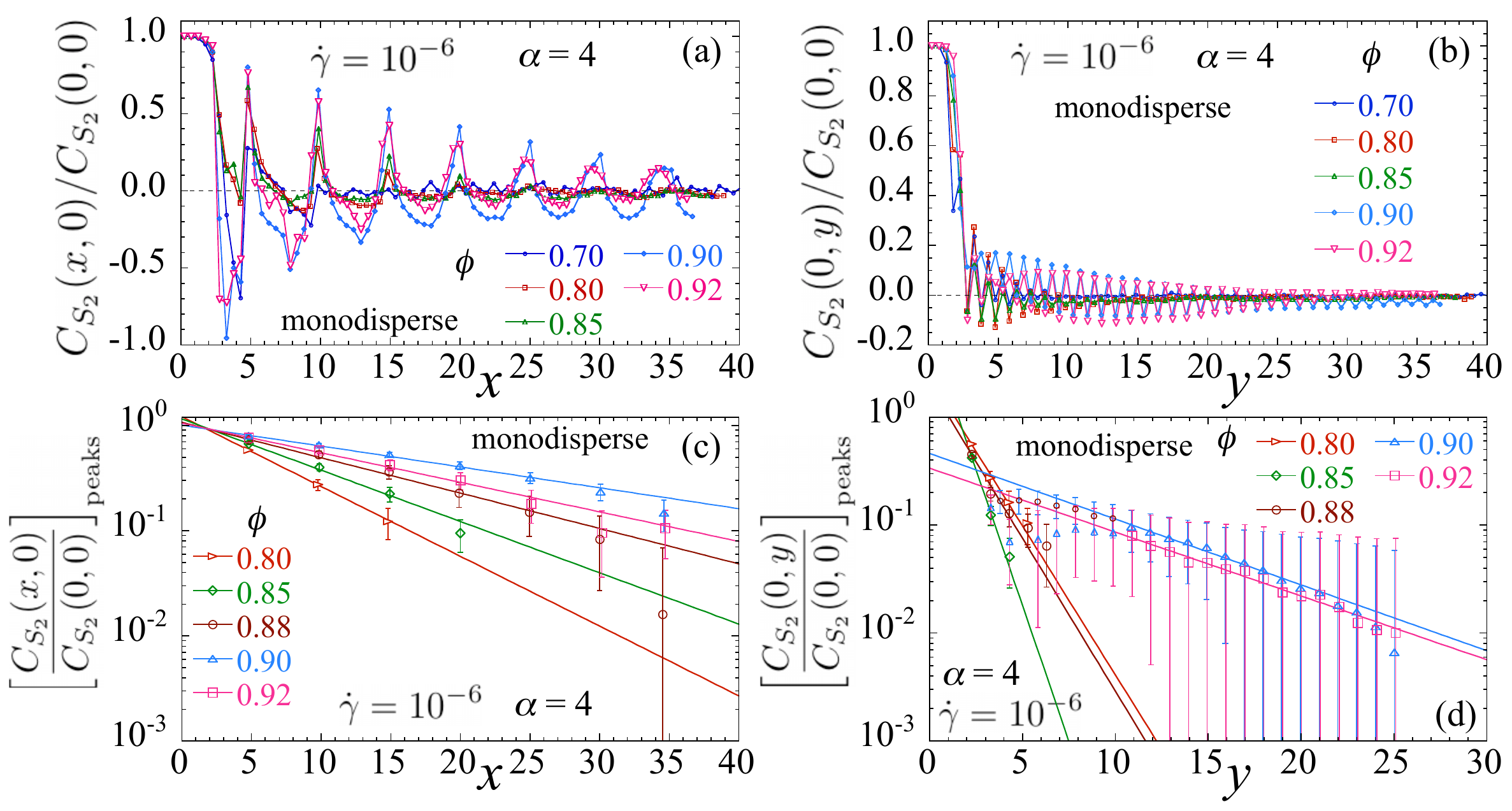}
\caption{For $N=1024$ monodisperse particles of asphericity $\alpha=4$, sheared at $\dot\gamma=10^{-6}$: Nematic order parameter correlation $C_{S_2}(\mathbf{r})/C_{S_2}(0)$ vs coordinates (a) $x$ and (b) $y$, parallel and perpendicular to the flow direction $\mathbf{\hat x}$, for different packing fractions $\phi$. Peak heights in (c) $C_{S_2}(x,0)/C_{S_2}(0,0)$ vs $x$ for $x_m\approx 5m$, and in (d) $C_{S_2}(0,y)/C_{S_2}(0,0)$ vs $y$ for $y_m\approx m$; straight lines are fits to an exponential decay.  Lengths are measured in units of the particle diameter, $2R=1$ and the system width is $L\approx 90$.
}
\label{CS2-mono} 
\end{figure}

We now consider the correlations of the nematic order parameter, computing $C_{S_2}(\mathbf{r})$ as defined earlier in Sec.~\ref{sNC}.
In Figs.~\ref{CS2-mono}(a) and \ref{CS2-mono}(b) we show  plots of  $C_{S_2}(\mathbf{r})/C_{S_2}(0)$ vs $x$ and $y$, parallel and perpendicular to the flow direction.  Comparing to Figs.~\ref{CS2}(c) and \ref{CS2}(d) for a bidisperse system, we see that the peaks in the $x$ direction are again sharper, with periodicity of $\Delta x=5$,  and persist to longer length scales.  Along the $y$ direction we see sharp oscillations with periodicity $\Delta y=1$, but the heights decay more quickly.  In Figs.~\ref{CS2-mono}(c) and \ref{CS2-mono}(d) we plot the peak heights vs $x$ and $y$ and fit to an exponential decay.  For the smaller $\phi=0.70$ and 0.75 the peak heights decay too quickly for an accurate determination, and we omit these from Figs.~\ref{CS2-mono}(c) and \ref{CS2-mono}(d).  For the $y$ direction, shown in Fig.~\ref{CS2-mono}(d) the heights can be non-monotonic, and the location of the peaks varies somewhat with $\phi$;  errors are large and so our fits should be regarded as just estimates.

\begin{figure}
\centering
\includegraphics[width=3.5in]{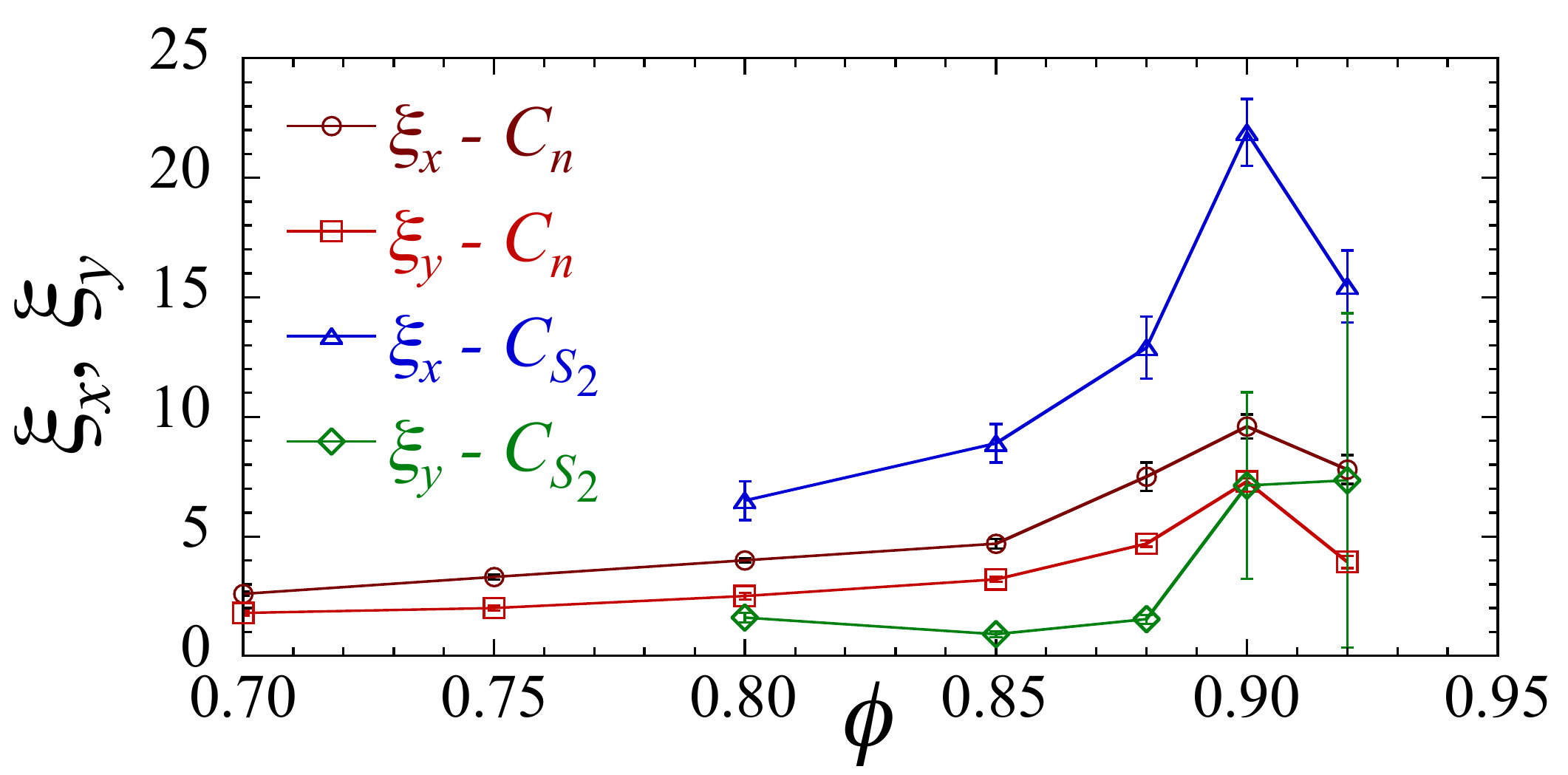}
\caption{For $N=1024$ monodisperse particles of asphericity $\alpha=4$, sheared at $\dot\gamma=10^{-6}$: Correlation lengths in the $x$ and $y$ directions, parallel and perpendicular to the flow, as obtained from the exponential fits to the peaks in the density correlation $C_n(\mathbf{r})$ and the nematic order parameter correlation $C_{S_2}(\mathbf{r})$, shown in Figs.~\ref{Cn-mono}(c) and \ref{Cn-mono}(d) and \ref{CS2-mono}(c) and \ref{CS2-mono}(d).
}
\label{xi-mono} 
\end{figure}

In Fig.~\ref{xi-mono} we show the decay lengths $\xi_x$ and $\xi_y$ that come from the exponential fits of Figs.~\ref{Cn-mono}(c) and \ref{Cn-mono}(d) for the positional correlation $C_n(\mathbf{r})$, and from Figs.~\ref{CS2-mono}(c) and \ref{CS2-mono}(d) for the nematic correlation $C_{S_2}(\mathbf{r})$.  From the positional correlation  $C_n$ we get a decay length in the $x$ direction that varies between 2.6 and 9.6 over the range of $\phi$ shown; in the $y$ direction the decay length varies between 1.8 and 7.3.  These are roughly twice as large as the corresponding decay lengths for the bidisperse system, but still no greater than two particle lengths.  The monodisperse system thus does not have any long range translational order.  From the nematic order parameter correlation $C_{S_2}$ we get a decay length in the $y$ direction that varies between 1.5 and 7.3, comparable to that found from $C_n$.  In the $x$ direction the decay length from $C_{S_2}$ varies between 6.5 and 22, roughly double that found from $C_n$.  The largest value $\xi_x\approx 22\approx L/4$ is roughly one quarter the length of the system, and so in Fig.~\ref{CS2-mono}(a) one does not see the peaks in $C_{S_2}(x,0)$ decaying to zero, although from Fig.~\ref{CS2-mono}(c) the decay does appear to be exponential.  Simulations of a larger length system  would be needed to confirm that the value $\xi_x\approx 22$ really is finite, and that there is no long range nematic ordering.

\subsection{Global Nematic Ordering}
\label{GNO}

\begin{figure}
\centering
\includegraphics[width=3.5in]{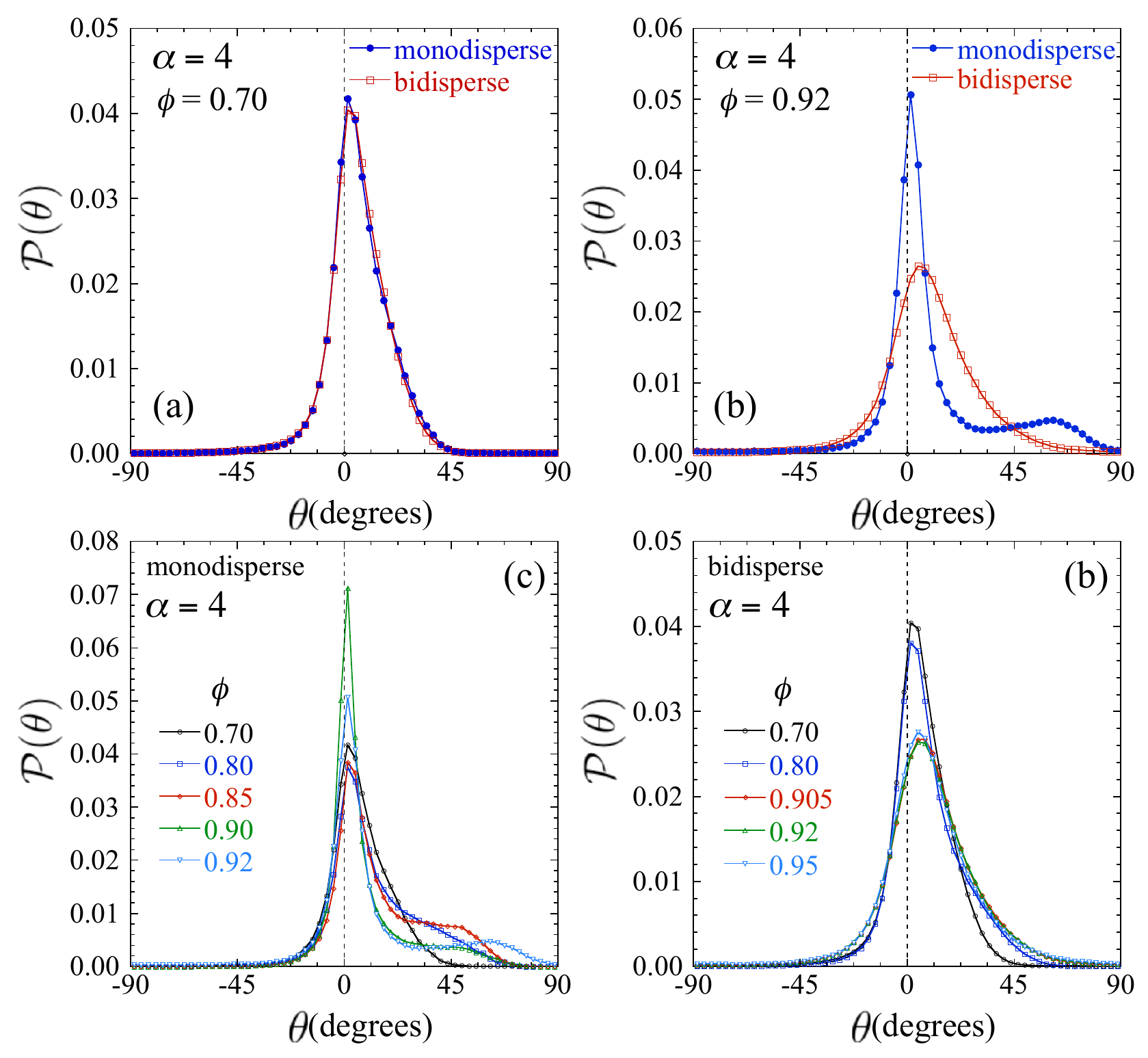}
\caption{Probability density $\mathcal{P}(\theta)$ for particles of asphericity $\alpha=4$ to be oriented at angle $\theta$ with respect to the flow direction: comparing size-monodisperse and size-bidisperse particles at packings (a) $\phi=0.70$ and (b) $\phi=0.92$; (c) monodisperse particles at different $\phi$, and (d) bidisperse particles at different  $\phi$.  Strain rate is $\dot\gamma=10^{-6}$ for the monodisperse system and $\dot\gamma=10^{-5}$ for the bidisperse system.
}
\label{Ptheta} 
\end{figure}

Finally, we consider several quantities related to the global nematic ordering of the system.  We are interested in how the differing packing geometry and greater local spatial ordering found in dense packings of the monodisperse system, as compared to the bidisperse system, will effect such orientational ordering.

We start by returning to an observation made at the start of this Sec.~\ref{sec:Monodisperse} on monodisperse systems.  Looking at the dense monodisperse configuration of Fig.~\ref{mono-config}, the particles generally appear to be more aligned with the flow directions as compared with the dense bidisperse configuration of Fig.~\ref{configs}(a).  To quantify this observation,  in Fig.~\ref{Ptheta} we plot the probability density $\mathcal{P}(\theta)$ for a particle to be oriented at angle $\theta$.  In Fig.~\ref{Ptheta}(a) we compare $\mathcal{P}(\theta)$ for monodisperse and bidisperse systems, both with particles of asphericity $\alpha=4$, at the relatively dilute packing $\phi=0.70$.  Here, we see that the distributions for the two cases are essentially identical.  However, in Fig.~\ref{Ptheta}(b) we compare the two cases at the denser packing $\phi=0.92$.  Here, we see a rather dramatic difference.  For the bidisperse case, $\mathcal{P}(\theta)$ is qualitatively similar to that at the lower packing, with a broad unimodal peak that is skewed to the right.  For the monodisperse case, however, we see a primary peak that remains comparatively sharp and centered close to zero at $\theta_\mathrm{peak}\approx 1.5^\circ$, but there is also a shoulder extending to larger angles that becomes a smaller secondary peak around $\theta\approx 60^\circ$.   
In Fig.~\ref{Ptheta}(c) we plot $\mathcal{P}(\theta)$ for the monodisperse system for several different packings from $\phi=0.70$ to 0.92, to show how this secondary peak develops as $\phi$ increases.  In Fig.~\ref{Ptheta}(d) we similarly plot $\mathcal{P}(\theta)$ at different $\phi$ for the bidisperse case;  we see that the width of the distribution broadens and the location of the peak shifts to slightly larger $\theta$ as $\phi$ increases, but otherwise the shape of the distribution stays qualitatively the same.

%Thus in both cases, as we will see explicitly later in Fig.~\ref{mono}(c), the direction $\theta_2$ of the nematic order parameter (given by the average of $\mathcal{P}(\theta)\cos2\theta$) increases as $\phi$ increases; however, unlike the bidisperse case, we find for the monodisperse case that the most probable orientation remains close to $\theta=0$.  We now consider other quantitative measures of the spatial structure of the monodisperse system, which will show further notable differences compared to the bidisperse case.

\begin{figure}
\centering
\includegraphics[width=3.5in]{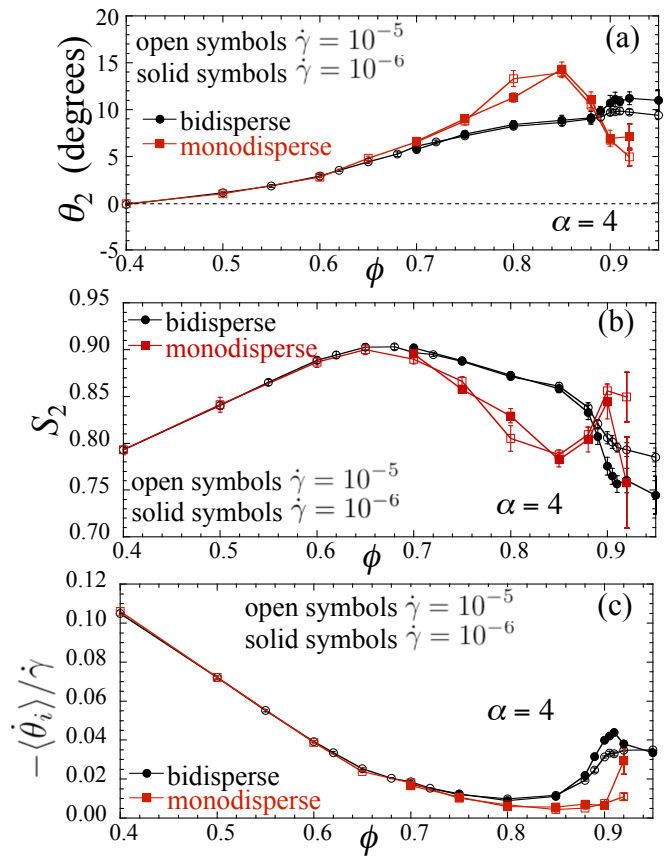}
\caption{(a) Orientation of the nematic order parameter $\theta_2$, (b) magnitude of the nematic order parameter $S_2$, and (c) average angular velocity $-\langle\dot\theta_i\rangle/\dot\gamma$,  vs packing $\phi$, for size monodisperse compared to size bidisperse spherocylinders of asphericity $\alpha=4$.  Open symbols are for a strain rate $\dot\gamma=10^{-5}$ while solid symbols are for $\dot\gamma=10^{-6}$.  For the bidisperse system $\phi_J\approx0.906$; for the monodisperse system we estimate $\phi_J\approx 0.92$.
}
\label{mono} 
\end{figure}

While the distributions $\mathcal{P}(\theta)$ for monodisperse and bidisperse systems are thus significantly different for dense packings, it is interesting to consider a measure of the average particle orientation.  This is most naturally given by the orientation $\theta_2$ of the nematic order parameter, which is computed from the individual particle orientations by Eq.~(\ref{eS2g2}); the sums in that equation are equivalent to averages over the distribution $\mathcal{P}(\theta)$.  In Fig.~\ref{mono}(a) we plot the resulting $\theta_2$ vs packing $\phi$ for the monodisperse system considered in this section, as compared to the bidisperse system studied in Sec.~\ref{sec:Bidisperse}.  We show results for the two strain rates $\dot\gamma=10^{-5}$ (open symbols) and $\dot\gamma=10^{-6}$ (solid symbols).  
Just as we saw in Fig.~\ref{Ptheta}(a) that $\mathcal{P}(\theta)$ was the same for monodisperse and bidisperse systems at low $\phi$, here  we see that $\theta_2$ for the two cases are similarly equal at low $\phi$.  However, as $\phi$ increases and the distributions $\mathcal{P}(\theta)$ start to differ, so do the values of $\theta_2$ for the two cases differ, though in both cases $\theta_2$ remains in the range $5-10^\circ$.  It is interesting to note that, for some range of $\phi$, the value of $\theta_2$ for the monodisperse system is greater than that for the bidisperse system, even though the monodisperse $\mathcal{P}(\theta)$ has a sharper peak that lies closer to $\theta\approx 0$.  This is presumably due to the weight in the broad shoulder that extends to larger angles.

In Figs.~\ref{mono}(b) and \ref{mono}(c) we show similar comparisons between monodisperse and bidisperse systems for  
the magnitude of the global nematic order parameter $S_2$, and the average particle angular velocity $-\langle\dot\theta_i\rangle/\dot\gamma$, respectively.  As with $\theta_2$ we see that these quantities agree between the monodisperse and bidisperse systems for low $\phi\lesssim 0.70$, but they differ for denser packings.

 In a previous work \cite{MT2}, that dealt strictly with bidisperse systems,  we  argued that the peak in the nonmonotonic $S_2$ marks a crossover from a region  of qualitatively single particle behavior below  $\phi_{S_2\,\mathrm{max}}$, to a region above $\phi_{S_2\,\mathrm{max}}$ 
%where decreasing free volume inhibits particle rotations; rotations occur at random uncorrelated times when fluctuations in the particle configuration open a local region of greater free volume that permits a rotation to take place.  Thus behavior above  $\phi_{S_2\,\mathrm{max}}$ 
where decreasing free volume causes behavior to be dominated by the local structure of the dense packing. 
The results in Figs.~\ref{Ptheta} and \ref{mono} give strong support for this scenario.  At small $\phi\lesssim\phi_{S_2\,\mathrm{max}}$ we see that $\mathcal{P}(\theta)$, $\theta_2$, $S_2$, and  $-\langle\dot\theta_i\rangle/\dot\gamma$ are essentially equal for the monodisperse and the bidisperse systems.  This is as would be expected for a single-particle-like limit, where the size of the particle would play no role in determining these quantities \cite{MT2}.
However at larger $\phi$, the results in Figs.~\ref{Cn} and \ref{Cn-mono} for positional correlations, and in Figs.~\ref{CS2} and \ref{CS2-mono} for nematic correlations, show that the monodisperse system has a much stronger local order than the bidisperse system.  The differences we find in $\mathcal{P}(\theta)$, $\theta_2$, $S_2$, and  $-\langle\dot\theta_i\rangle/\dot\gamma$ in such dense packings above $\phi_{S_2\,\mathrm{max}}$ thus reflect this difference in local packing structure.

%In particular, both  $-\langle\dot\theta_i\rangle/\dot\gamma$ and $S_2$ for the monodisperse system become noticeably  smaller than for the bidisperse system, indicating that particles  rotate more slowly but are also less orientationally ordered.  
%However as $\phi$ increases further, approaching jamming, we see that $S_2$ for the monodisperse system starts to increase, suggesting that jamming helps order the system.  
%Since we know from our  results presented above that the structure of the dense packings above $\phi_{S_2\,\mathrm{max}}$ is noticeably different for monodisperse as compared to bidisperse systems, with the former showing local ordering to longer length scales, and with the  distribution of particle orientations $\mathcal{P}(\theta)$ being qualitatively different, it is therefore not surprising to see that $-\langle\dot\theta_i\rangle/\dot\gamma$ and $S_2$ for the two types of systems become noticeably different in this region of dense packing.

In our prior work \cite{MT2} we discussed how the orientation of particles appears to arise from a   competition between aligning with the shear flow, as an isolated particle would do, vs aligning with the direction of minimal stress.  The details of this remain poorly understood.  It would appear that the strong local ordering of the monodisperse system at dense packings, as indicated by Figs.~\ref{Cn-mono} and \ref{CS2-mono}, shifts this competition to favor increased alignment of many of the particles parallel to the flow.

%%%%%%%%%%%%%%%%%%%%%

\section{Shearing  Highly Ordered Configurations}
\label{sec:HOR}

In the previous parts of this work, as well as in our earlier works \cite{MT1,MT2}, we began our shearing simulations from a random initial configuration, and shear to large total strains $\gamma$ so as to reach the steady state.  The assumption, motivated by results for sheared circular disks \cite{Vagberg.PRE.2011}, is that by shearing long enough, one creates a well defined ensemble of states that is independent of the initial configuration.  In contrast, one can wonder whether the same steady-state ensemble will result if one starts from an initial configuration of locally well ordered particles.  Will such a system remain ordered as it shears, or will it revert to the same ensemble obtained from the random initial configurations?
In this section we investigate this question for spherocylinders of asphericity $\alpha=4$.
We consider, for systems of both size-bidisperse and size-monodisperse particles, several different initial configurations designed to be locally ordered in such a way that we can pack particles to large density without any particle overlaps.  

\subsection{Size-Bidisperse Particles}

\begin{figure}
\centering
\includegraphics[width=3.5in]{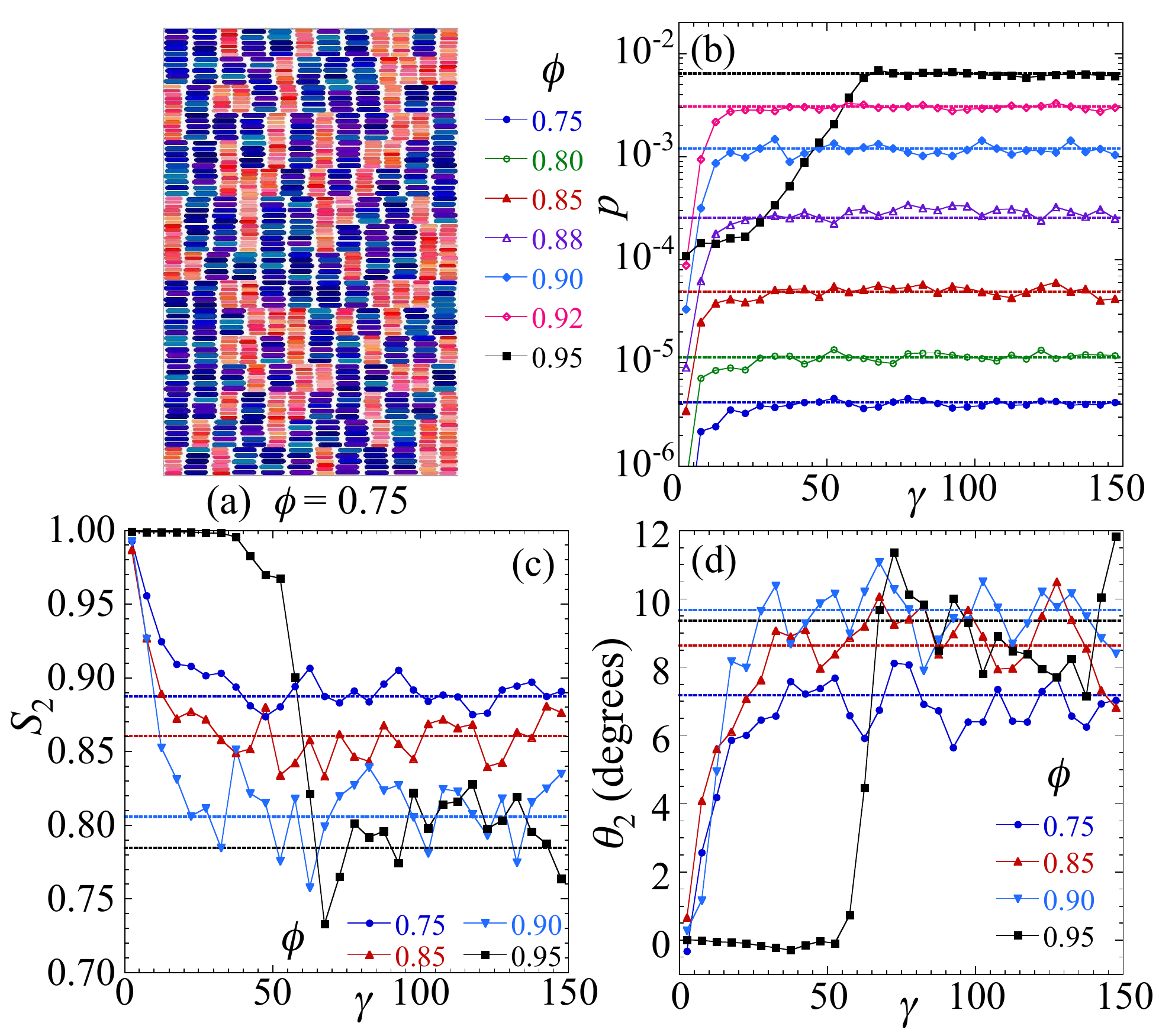}
\caption{(a) Snapshot of a size bidisperse configuration of locally ordered stacks of particles at a packing $\phi=0.75$; big spherocylinders are shown in blue hues, while small spherocylinders are shown in red hues.  Shearing initial configurations as in (a) at the strain rate $\dot\gamma=10^{-5}$, we show (b) pressure $p$, and (c) magnitude $S_2$ and (d) orientation $\theta_2$ of the nematic order parameter vs net strain $\gamma$ at different packings $\phi$.  The data points in (b), (c), and (d)  represent averages of the instantaneous values over  strain windows of $\Delta\gamma=5$.  The dotted horizontal lines in (b), (c) and (d) give the ensemble averaged values when starting from a random initial configuration.  A reduced set of $\phi$ are shown in (c) and (d) for clarity.  Animations of the shearing at $\phi=0.90$ and 0.95 are available in our Supplemental Material \cite{SM}.
}
\label{bi-ord-i} 
\end{figure}

We start by constructing a close packed, locally ordered, configuration as follows.    We take a stack of five big spherocylinders, all oriented parallel to the flow direction $\mathbf{\hat x}$ and lying perfectly one on top of another so that their centers of mass align vertically.  We then take a stack of seven small spherocylinders in the same fashion; the heights of these two stacks are equal (recall, $R_b/R_s=1.4=7/5$).  We then  randomly place seven stacks of the big particles and five stacks of the small particles next to each other in a horizontal row, so that there are the same number of big and small particles in this row of stacks.  We then construct 16 such rows of stacks, each row being constructed in an independent random fashion, so that we have a total of $N=1120$ particles.
We then affinely expand the system to the desired packing fraction $\phi$, and introduce a small length scale disorder by making a random displacement of each particle, with the displacement sampled uniformly over the particle's Voronoi cell.  The resulting configuration contains no particle overlaps.  An example of such an initial configuration at the packing $\phi=0.75$ is shown in Fig.~\ref{bi-ord-i}(a).  In this and subsequent similar figures, blue hues are used for the big particles and red hues for the small particles, but in each case we use a small spread of colors so as to help distinguish different particles.

Shearing such initial, locally ordered, configurations at a strain rate $\dot\gamma=10^{-5}$ we compute the instantaneous pressure $p(\gamma)$, as well as the magnitude $S_2(\gamma)$ and orientation $\theta_2(\gamma)$ of the nematic order parameter.  Because   fluctuations in these instantaneous values can be large, we choose to smooth out the fluctuations by averaging the instantaneous values over a strain window of width $\Delta\gamma=5$.  We plot the resulting strain averaged values of $p$, $S_2$ and $\theta_2$ in Figs.~\ref{bi-ord-i}(b), \ref{bi-ord-i}(c), and \ref{bi-ord-i}(d), respectively, for a range of packings $\phi$.  The dotted horizontal lines in these figures  give the ensemble averaged values of these quantities, when starting from a random initial configuration, as obtained from our earlier work in Refs.~\cite{MT1,MT2}.

For all $\phi$ we see that $p$ starts from  zero in the initial configuration with no particle overlaps, but then rises to saturate at the same value as obtained from a random initial configuration.  Similarly, the nematic order parameter starts from an initial $S_2=1$ and $\theta_2=0$, but then evolves to saturate at the same values of $S_2$ and $\theta_2$ found when shearing from a random initial configuration.
Shearing an initial, locally ordered, size-bidisperse configuration constructed as in  Fig.~\ref{bi-ord-i}(a) thus results in the same spatially disordered steady-state ensemble as obtained from an initial random configuration.  This disordering is readily seen in animations of the shearing at   $\phi=0.90$ and 0.95,  which are available in our Supplemental Material \cite{SM}.  From Figs.~\ref{bi-ord-i}(b), \ref{bi-ord-i}(c) and \ref{bi-ord-i}(d) we see that this disordering takes place fairly quickly,  except for $\phi=0.95$ which is considerably above the jamming $\phi_J=0.906$; in that latter case the system stays ordered up to some considerable strain $\gamma\approx 60$, but then disorders just as at the smaller $\phi$. 

\begin{figure}
\centering
\includegraphics[width=3.5in]{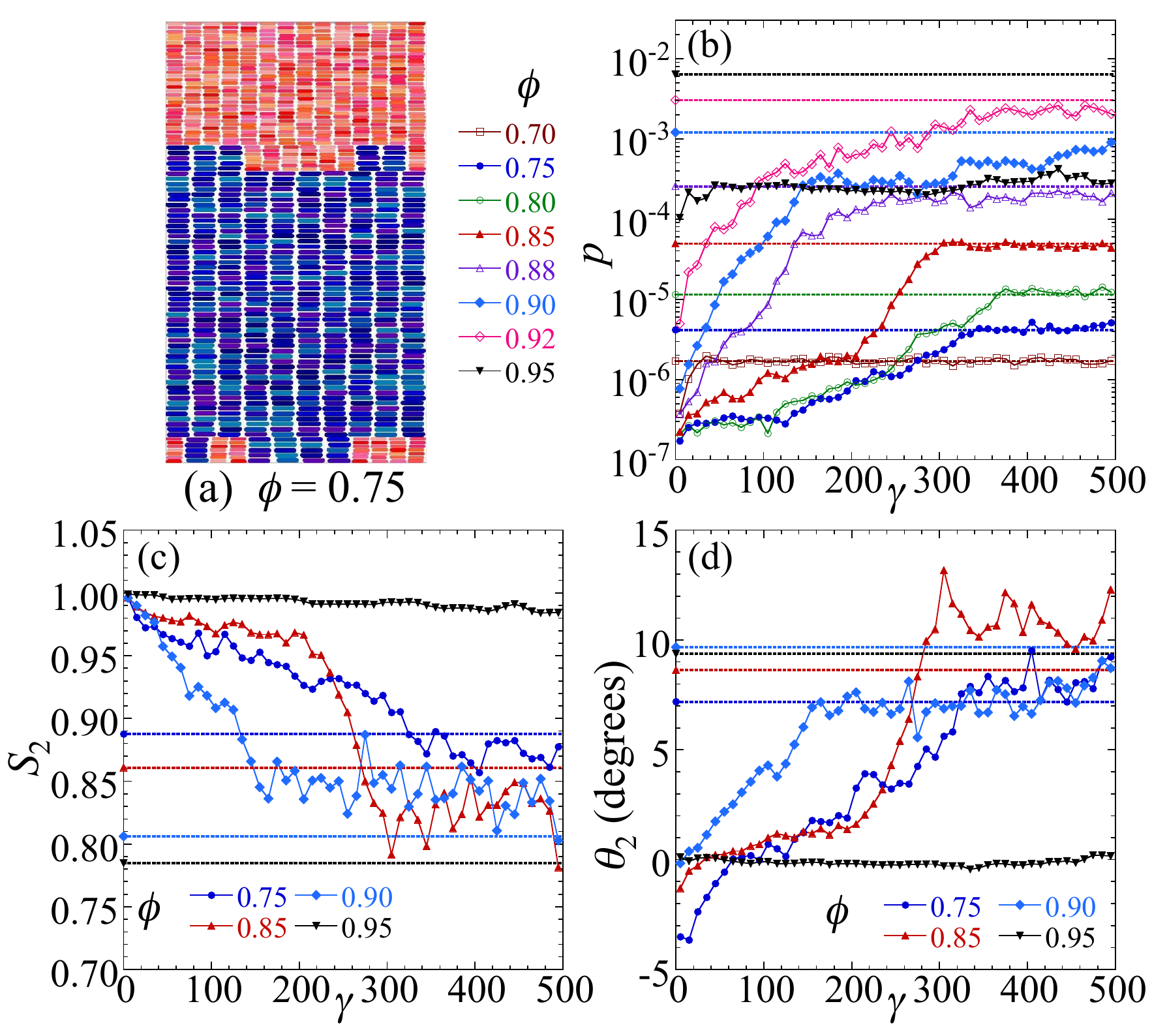}
\caption{(a) Snapshot of a size bidisperse, phase separated, configuration of particles at a packing $\phi=0.75$; big spherocylinders are shown in blue hues, while small spherocylinders are shown in red hues.  The interfaces between the regions of big and small particles consist of a random set of stacks of five big particles and seven small particles.  Shearing initial configurations as in (a) at the strain rate $\dot\gamma=10^{-5}$, we show (b) pressure $p$, and (c) magnitude $S_2$ and (d) orientation $\theta_2$ of the nematic order parameter vs net strain $\gamma$ at different packings $\phi$.  The data points in (b), (c), and (d)  represent averages of the instantaneous values over  strain windows of $\Delta\gamma=10$.  The dotted horizontal lines in (b), (c) and (d) give the ensemble averaged values when starting from a random initial configuration.  A reduced set of $\phi$ are shown in (c) and (d) for clarity.  Animations of the shearing at $\phi=0.90$ and 0.95 are available in our Supplemental Material \cite{SM}.
}
\label{bi-ord-ii} 
\end{figure}

We next consider an initial configuration that is even more ordered than that of Fig.~\ref{bi-ord-i}(a).   We start with stacks of ordered big and small spherocylinders as described above, but now we phase separate the particles so that the big particles are all on the bottom of the system while the small particles are all on the top of the system.  At each of the two horizontal interfaces between big and small particles (there are two interfaces due to our periodic Lees-Edwards boundary conditions) we put a randomly ordered row consisting of seven stacks of five big particles and five stacks of seven small particles, as in the case previously discussed.  
We then affinely expand the system to the desired packing fraction $\phi$, and make a random displacement of each particle uniformly over its Voronoi cell, so that the resulting configuration has no particle overlaps.  An example of such an initial configuration at the packing $\phi = 0.75$ is shown in Fig.~\ref{bi-ord-ii}(a).

Shearing such configurations at a strain rate $\dot\gamma=10^{-5}$, in Figs.~\ref{bi-ord-ii}(b), \ref{bi-ord-ii}(c), and \ref{bi-ord-ii}(d) we plot the resulting $p$, $S_2$, and $\theta_2$ vs $\gamma$, obtained by averaging over strain windows of $\Delta\gamma=10$, for a range of packings $\phi$.  We see from Fig.~\ref{bi-ord-ii}(b) that for all packings, except the largest $\phi=0.95$, the pressure $p$ increases and appears to saturate at the same value found for the ensemble average starting from a random initial configuration.  This suggests that the phase separated initial configurations are disordering as they are sheared.  However, considering Figs.~\ref{bi-ord-ii}(c) and \ref{bi-ord-ii}(d), it is less clear whether $S_2$ and $\theta_2$ are saturating to the same values as when shearing from a random initial configuration.  

\begin{figure}
\centering
\includegraphics[width=3.in]{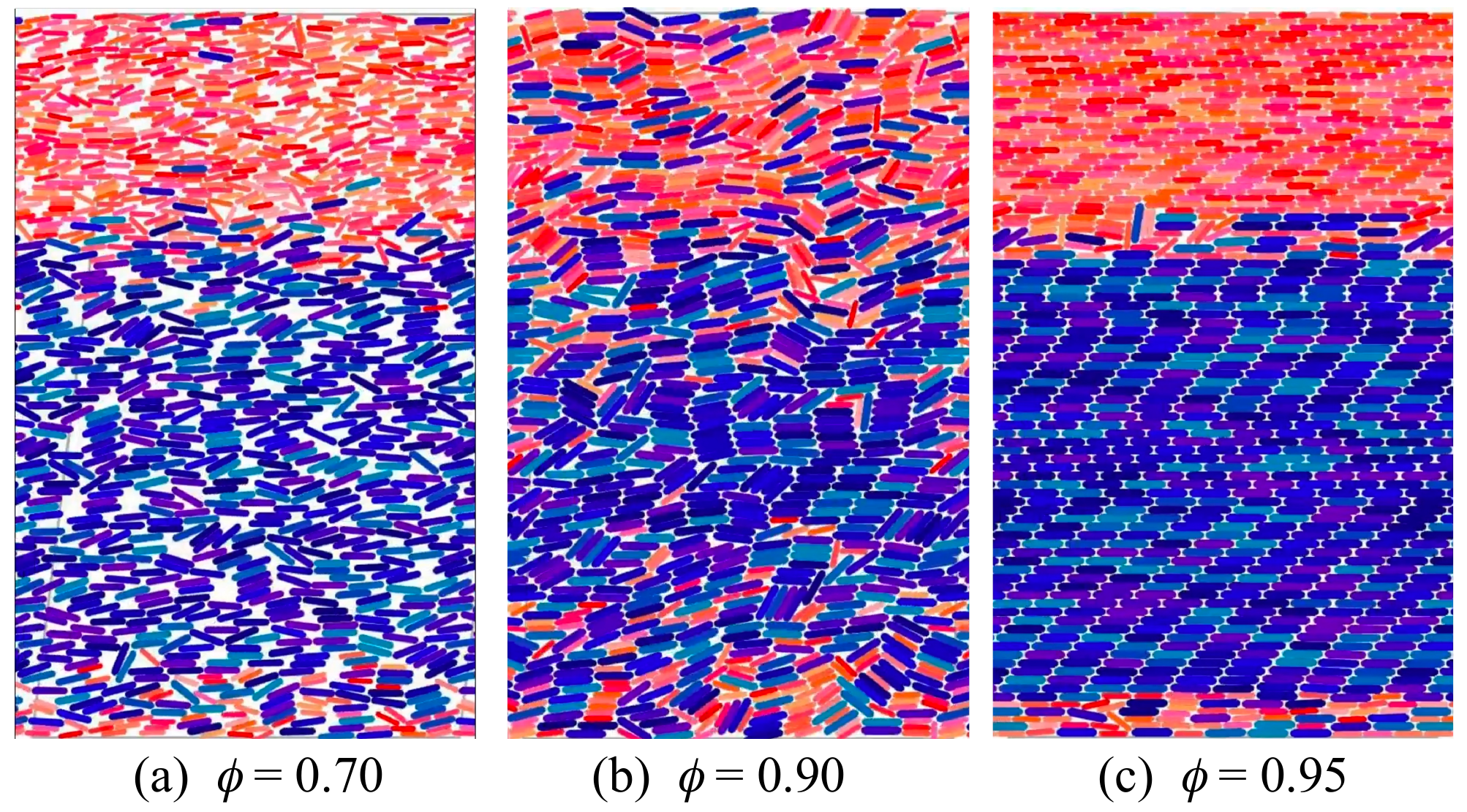}
\caption{Snapshots of the final configurations, after a total shear strain of $\gamma=500$, corresponding to initial phase separated configurations as in Fig.~\ref{bi-ord-ii}(a).  Results are shown for packings (a) $\phi=0.70$, (b) $\phi=0.90$, and (c) $\phi=0.95$, sheared at a rate $\dot\gamma=10^{-5}$.  Animations of the shearing at $\phi=0.90$ and 0.95 are available in our Supplemental Material \cite{SM}.
}
\label{bi-ord-ii-end} 
\end{figure}

To see what is happening, in Fig.~\ref{bi-ord-ii-end} we show snapshots of the final configurations obtained after shearing the initial configurations as in Fig.~\ref{bi-ord-ii}(a) to a total shear strain $\gamma=500$.  While the system at $\phi=0.95$, shown in Fig.~\ref{bi-ord-ii-end}(c), stays mostly phase separated and highly orientationally  ordered, 
we see that for $\phi=0.70$ and 0.90, shown in Figs.~\ref{bi-ord-ii}(a) and \ref{bi-ord-ii}(b), the system remains phase separated to a considerable degree, but each of the regions of big and small particles has separately decreased its orientational ordering.  
Because the values of $S_2$ and $\theta_2$ are different comparing bidisperse and monodisperse systems, as shown in Fig.~\ref{mono}, it is thus not surprising that the $S_2$ and $\theta_2$ that we find here for our phase separated system is not quite in agreement with what is found when shearing from a bidisperse random initial configuration.

\begin{figure}
\centering
\includegraphics[width=3.5in]{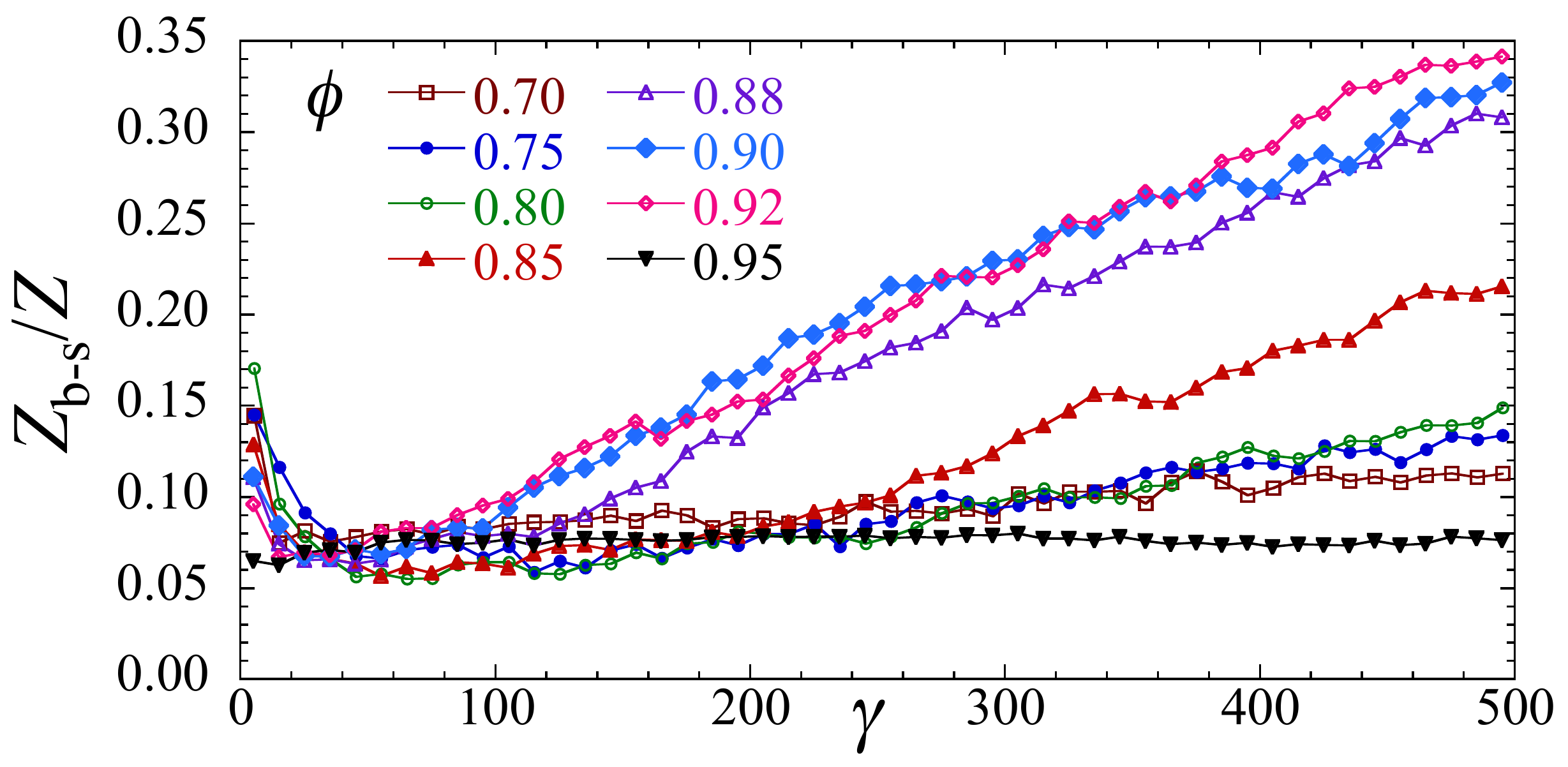}
\caption{For initial configurations as in Fig.~\ref{bi-ord-ii}(a), the ratio of the number of contacts per particle between big and small spherocylinders, $Z_\mathrm{b\text{-}s}$, to the total number of all contacts per particle, $Z$, vs shear strain $\gamma$ for systems at different packings $\phi$.  Data points represent averages of the instantaneous values over strain windows of $\Delta\gamma=10$.  The system is sheared at a rate $\dot\gamma=10^{-5}$.
}
\label{Zbs} 
\end{figure}

Comparing the configurations shown in Figs.~\ref{bi-ord-ii-end}(a) and \ref{bi-ord-ii-end}(b), we see that the width of the interface between the two regions, and the penetration of one phase into the other, seems to increase as the packing $\phi$ increases.   
To quantify this  observation, we compute the following.  If $Z$ is the average number of contacts per particle, and $Z_\mathrm{b\text{-}s}$ is the average number of contacts between big and small particles per particle, in Fig.~\ref{Zbs} we plot the ratio $Z_\mathrm{b\text{-}s}/Z$ vs strain $\gamma$ at different packings $\phi$.  Each data point in Fig.~\ref{Zbs} is an average of the instantaneous $Z_\mathrm{b\text{-}s}/Z$ over a strain window of $\Delta\gamma=10$.  The larger is the fraction $Z_\mathrm{b\text{-}s}/Z$, the more contacts there are between big and small particles, and the less is the extent of the phase separation.  
When shearing from a random initial configuration one finds in the steady-state that $Z_\mathrm{b\text{-}s}/Z\approx 0.5$ at any packing $\phi$.
We clearly see in Fig.~\ref{Zbs} that, aside from an initial decrease at small strains $\gamma$, the ratio $Z_\mathrm{b\text{-}s}/Z$ steadily increases with increasing strain $\gamma$, suggesting that the big and small particles will completely mix if we are able to shear to large enough strains.
Moreover, as suggested 
by Figs.~\ref{bi-ord-ii-end}(a) and \ref{bi-ord-ii-end}(b), we see that $Z_\mathrm{b\text{-}s}/Z$ generally increases as $\phi$ increases, indicating a greater degree of phase mixing as the system gets denser.  The only exception is for the largest packing $\phi=0.95$ where $Z_\mathrm{b\text{-}s}/Z$ stays small and is constant with $\gamma$, indicating the persistence of the phase separated state in this dense packing.

We can understand the variation of $Z_\mathrm{b\text{-}s}/Z$ with the net strain $\gamma$ as follows.  The initial decrease at small $\gamma$ is because
in the initial configuration of non-overlapping particles there are no contacts of any type;  as the system first starts to shear, it is the particles within the interfaces between the regions of big and small particles that first come into contact, and so a large fraction of the particles that have any contacts at all have contacts with particles of a different size.  As shearing continues, however, particles in the bulk of the system form contacts as well; these are generally with particles of the same size, and so $Z_\mathrm{b\text{-}s}/Z$ decreases.  Finally, as the system shears further, the width of the interface region increases, and penetration of one phase into the other increases, so $Z_\mathrm{b\text{-}s}/Z$ now increases.  In this latter region  $Z_\mathrm{b\text{-}s}/Z$   steadily grows as $\gamma$ increases.  Animations of the shearing of these phase separated systems at $\phi=0.90$ and 0.95 are available in our Supplemental Material \cite{SM}.

As seen in Fig.~\ref{bi-ord-ii-end}, the shearing of the system both disorders the perfect orientational ordering of the initial configuration, as well as causes the big and small particles to mix.  The orientational disordering takes place on a faster strain scale than does the mixing.  The former may be estimated by the increase to saturation of the pressure in Fig.~\ref{bi-ord-ii}(b), and is presumably a result of  shear induced particle rotations.
The latter is measured by the behavior of $Z_\mathrm{b\text{-}s}/Z$ in Fig.~\ref{Zbs}, and is a result of the slower process of  transverse diffusion of particles at the interface.
It generally appears that both processes occur more rapidly as the packing $\phi$ increases.  We  speculate that the increased number and magnitude of collisions as $\phi$ increases leads to larger fluctuations and  thus a faster rate of disorienting and diffusing.  However the failure of $\phi=0.95$ to disorder indicates that this simple picture must be taken with caution.

We have also considered shearing from an initial configuration in which each row of particles is entirely composed of spherocylinders all of the same size.  Such rows of big or small spherocylinders are then stacked randomly.  We find that, for $\phi < 0.8$, such initial configurations disorder and result in the same steady-state ensemble as found from a random initial configuration.  For $\phi\ge 0.8$, however, the systems remain ordered at least up to the maximum strain $\gamma=200$ that we have simulated for these cases.

\subsection{Size-Monodisperse Particles}

For size-monodisperse systems we have already seen, in connection with Fig.~\ref{flow-mono}(c), that  at large packings the system can get locked into a spatially inhomogeneous flowing state, even when starting from an initial random configuration.  
Here, we consider what happens if the initial configuration is well ordered.  All our systems in this section are sheared at the rate $\dot\gamma=10^{-4}$.

\begin{figure}
\centering
\includegraphics[width=3.5in]{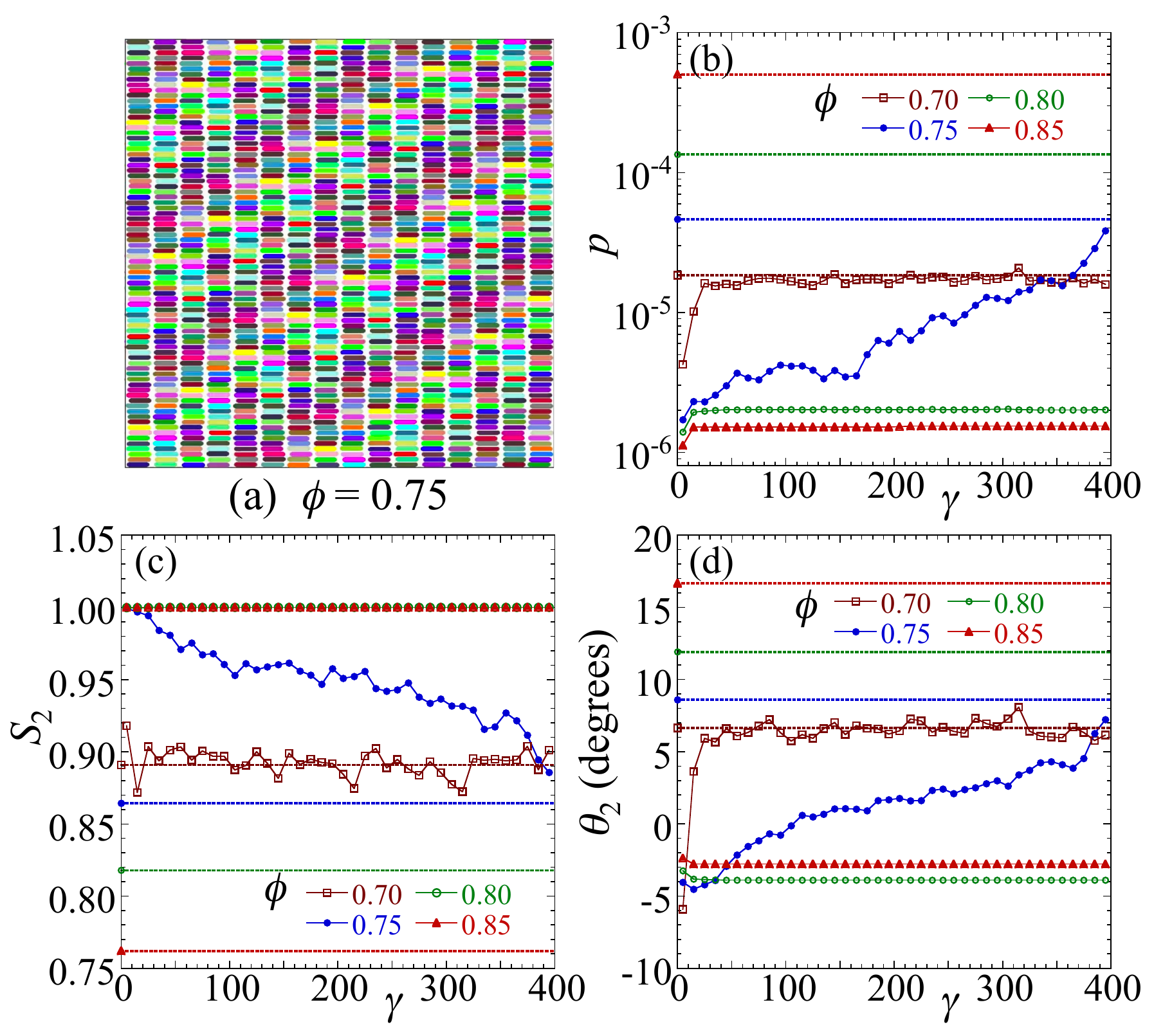}
\caption{(a) Snapshot of a size monodisperse, locally ordered, configuration of particles at a packing $\phi=0.75$; colors are used to help distinguish different particles and have no other meaning.  Shearing initial configurations as in (a) at the strain rate $\dot\gamma=10^{-4}$, we show (b) pressure $p$, and (c) magnitude $S_2$ and (d) orientation $\theta_2$ of the nematic order parameter vs net strain $\gamma$ at different packings $\phi$.  The data points in (b), (c), and (d)  represent averages of the instantaneous values over  strain windows of $\Delta\gamma=10$.  The dotted horizontal lines in (b), (c) and (d) give the ensemble averaged values when starting from a random initial configuration.  
Animations of the shearing at $\phi=0.75$ and 0.85 are available in our Supplemental Material \cite{SM}.
}
\label{mono-ord-i} 
\end{figure}

Since particles are monodisperse in size, it is easy to construct highly ordered configurations.  We start first with an ordered rectangular lattice of particles, all oriented along the flow direction $\mathbf{\hat x}$.  We then affinely expand the system to the desired packing fraction $\phi$, and introduce a small length scale disorder by making a random displacement of each particle uniformly over its Voronoi cell.  The resulting configuration has no particle overlaps.  An example of such an initial configuration at the packing $\phi=0.75$ is shown in Fig.~\ref{mono-ord-i}(a).

\begin{figure}
\centering
\includegraphics[width=3.5in]{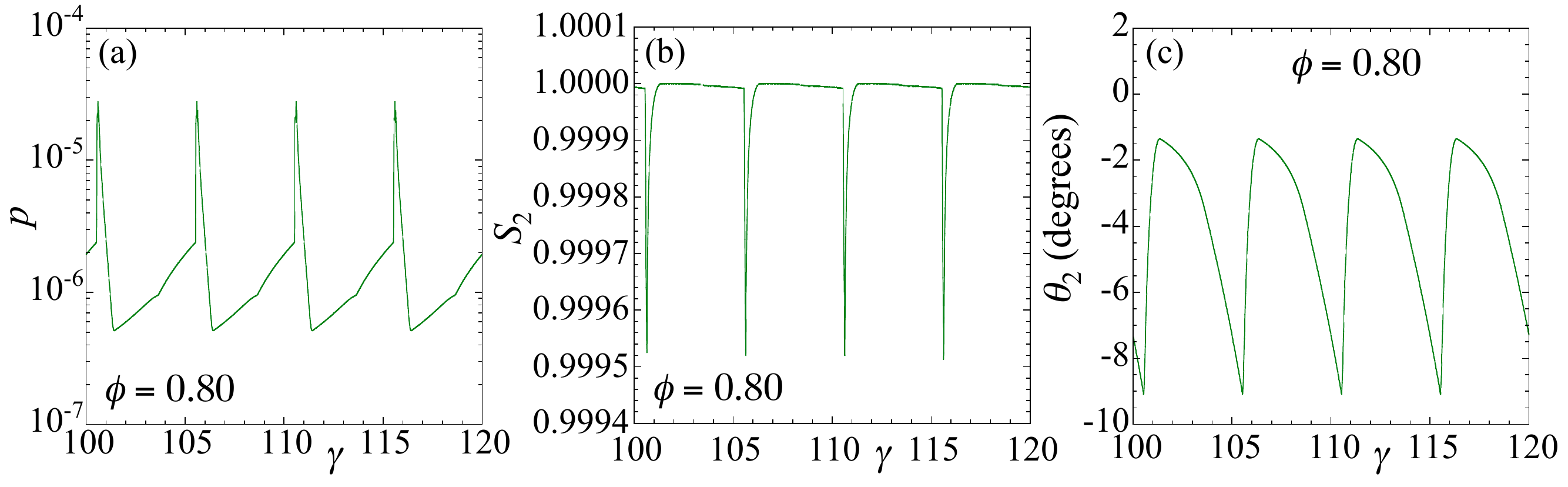}
\caption{Variation of the instantaneous (a) pressure $p$, and (b) magnitude $S_2$, and (c) orientation $\theta_2$ of the nematic order parameter, with shear strain $\gamma$, for the system of Fig.~\ref{mono-ord-i} at packing $\phi=0.80$.  The periodic behavior seen in these quantities illustrates the periodic wagging of the nematic order parameter in this highly ordered configuration.  The period of oscillation is $\gamma=5$, corresponding to the relative displacement of particles in adjacent rows by one particle length.
}
\label{mono-wag} 
\end{figure}

In Figs.~\ref{mono-ord-i}(b), \ref{mono-ord-i}(c), and \ref{mono-ord-i}(d) we show the resulting $p$, $S_2$, and $\theta_2$ vs $\gamma$ for a range of packings $\phi$, obtained by averaging the instantaneous values over strain windows of $\Delta\gamma=10$.  The dotted horizontal lines in these figures give the ensemble averaged values of these quantities when starting from a random initial configuration. The configuration at $\phi=0.70$ is seen to quickly disorder upon shearing,  reaching the same steady state as found from a random initial configuration.  At $\phi=0.75$ we see the system disordering, but over a much longer strain interval; only towards the end of our simulation, after a strain of $\gamma=400$, does it appear to be converging to the steady-state values found from a random initial configuration.  For $\phi=0.80$ and larger, the system remains in an ordered state for as long as we have sheared.  In such ordered states the particles show a periodic wagging of the nematic order parameter; the particles in a given row coherently rotate clockwise to negative angles $\theta_i<0$ below the flow direction, where they hit the particles in the row below them and then bounce back to start another cycle of oscillation.  This wagging is manifest in the periodic behavior of the instantaneous $p$, $S_2$, and $\theta_2$, as shown in Fig.~\ref{mono-wag} for the packing $\phi=0.80$.  The period of these oscillations is $\gamma=5$, corresponding to the relative displacement of particles in adjacent rows by one particle length.

\begin{figure}
\centering
\includegraphics[width=3.5in]{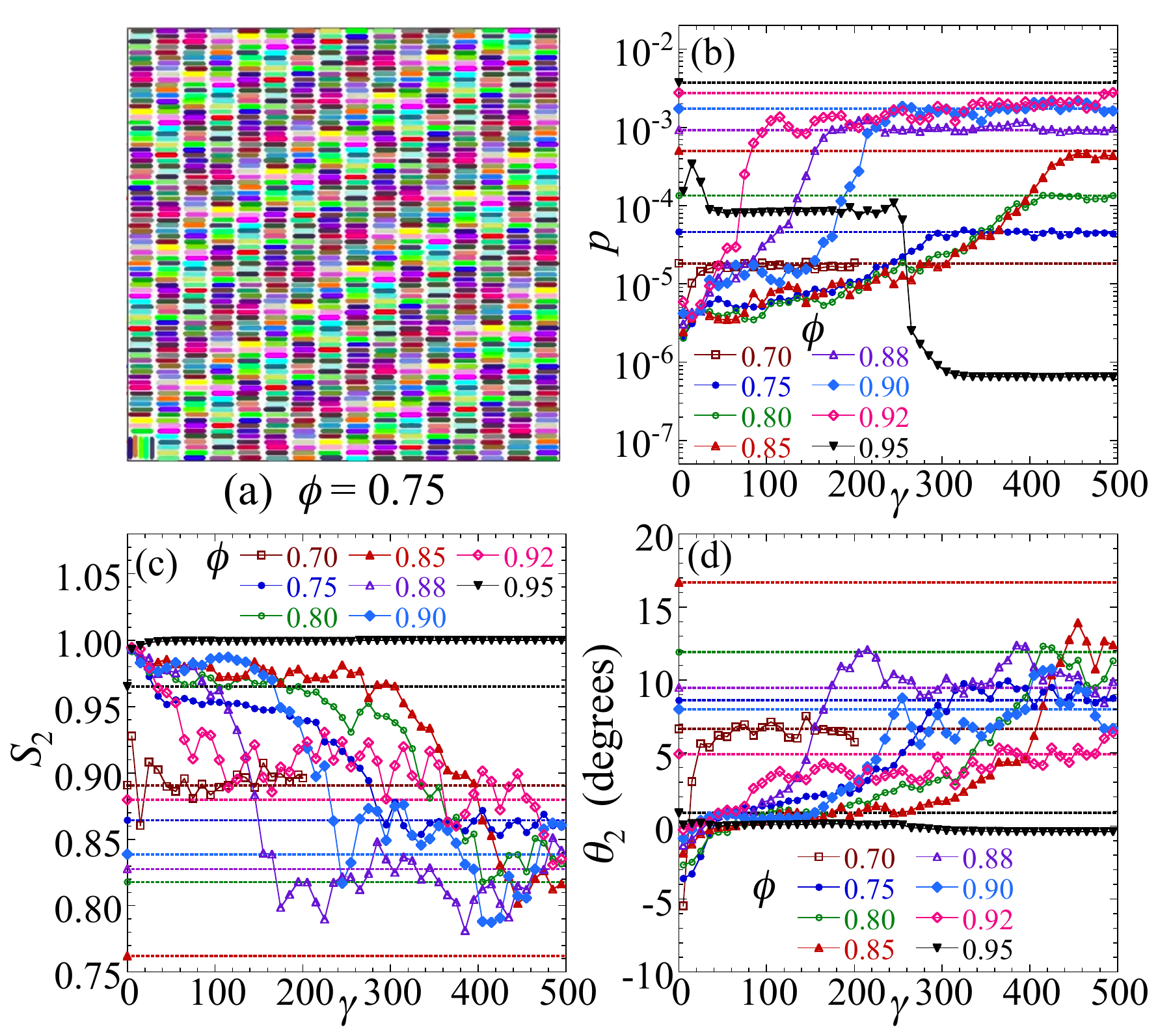}
\caption{(a) Snapshot of a size-monodisperse, locally ordered, configuration of particles at a packing $\phi=0.75$, where a defect has been introduced by the rotation by $90^\circ$ of five adjacent particles in the lower left corner; colors are used to help distinguish different particles and have no other meaning.  Shearing initial configurations as in (a) at the strain rate $\dot\gamma=10^{-4}$, we show (b) pressure $p$, and (c) magnitude $S_2$, and (d) orientation $\theta_2$ of the nematic order parameter vs net strain $\gamma$ at different packings $\phi$.  The data points in (b), (c), and (d)  represent averages of the instantaneous values over  strain windows of $\Delta\gamma=10$.  The dotted horizontal lines in (b), (c) and (d) give the ensemble averaged values when starting from a random initial configuration.  
Animations of the shearing at $\phi=0.90$ and 0.92 are available in our Supplemental Material \cite{SM}.
}
\label{mono-ord-ii} 
\end{figure}

To see how stable the ordered configurations of Fig.~\ref{mono-ord-i} are to preserving their order upon shearing at large density, we 
next  construct an initial configuration, starting just as before, but now introducing a new localized defect by rotating a group of five stacked particles by $90^\circ$, so that these are oriented perpendicular to the flow.  An example of such an initial configuration at the packing $\phi=0.75$ is shown in Fig.~\ref{mono-ord-ii}(a); the rotated particles are in the lower left corner of the image.  In Figs.~\ref{mono-ord-ii}(b), \ref{mono-ord-ii}(c), and \ref{mono-ord-ii}(d) we show the resulting $p$, $S_2$ and $\theta_2$ as such configurations are sheared at different packings $\phi$.   The plotted values are  obtained by averaging the instantaneous values over strain windows of $\Delta\gamma=10$.

In contrast to the behavior seen in Fig.~\ref{mono-ord-i}(b) for the defect free configuration, in Fig.~\ref{mono-ord-ii}(b) we see for all packings $\phi=0.70$ to $0.92$ that the system disorders as it shears, with the pressure rising from its initial small value to the same steady-state value found from a random initial configuration.  Interestingly, it is the larger $\phi$ that disorder more quickly than the smaller $\phi$.  In Figs. ~\ref{mono-ord-ii}(c) and \ref{mono-ord-ii}(d), although the data are more scattered, we see that $S_2$ and $\theta_2$ similarly reach the same values found from shearing from a random initial configuration; the only exception is for $\phi=0.85$ where $S_2$ remains larger and $\theta_2$ remains smaller, indicating that the initial configuration has not yet disordered to the extent found when shearing from a random initial configuration.  Looking at animations of the shearing, available in our Supplemental Material \cite{SM}, we see that the localized defect of  rotated particles, introduced in the initial configuration, induces a region of nearby disorder, that grows and eventually fills the system as the system is sheared.  For our larger packing $\phi=0.95$, however, we find that after a strain of $\gamma\approx 260$, the defect of rotated particles disappears, the particles all become aligned parallel to the flow, and the system persists in an ordered state for the remainder of the simulation up to $\gamma=500$.

We have also considered other particular initial configurations.  In one case we take the same configurations as in Fig.~\ref{mono-ord-i}(a) and then randomly displace the rows of particles in the horizontal direction, with all the particles in a given row displacing the same amount.  Such configurations behave qualitatively the same as the ones without the row displacements;  large packings $\phi$ remain ordered while small packings $\phi$ disorder, although the disordering takes place somewhat sooner and extends to a slightly larger $\phi$ than without the row displacements.  We have similarly taken the same configurations as in Fig.~\ref{mono-ord-i}(a) but then randomly displace the columns of particles in the vertical direction, with all the particles in a given column displacing the same amount.  In this case we find that all $\phi\le 0.88$ disorder by roughly $\gamma=50$, but larger $\phi\ge0.90$ remain ordered out to $\gamma=200$.

From our results in this section we conclude that, for both size-bidisperse and size-monodisperse systems, even highly ordered initial configurations will disorder upon shearing, and result in the same steady-state ensemble as found when starting from a random initial configuration, if the packing $\phi$ is small or moderate; for our spherocylinders with $\alpha=4$ we find this to be the case whenever $\phi<0.80$.  However, even for more dense systems, we find in many cases that the initial highly ordered configuration will also disorder and result in the same ensemble as found from a random initial configuration.  The initial configurations that remain highly ordered out to large total strains $\gamma$ seem to be those in which the particles are able to flow over each other in well defined channels, resulting only in a coherent wagging of the nematic order parameter.  However, when the initial configuration contains sufficient variation in the vertical alignment of particles, even if this occurs only locally, the wagging of particles near these vertical misalignments turns into full particle rotations, which then serve to increase and propagate  disorder in the flowing configuration.  We cannot rule out the possibility that even highly ordered initial configurations might eventually disorder if sheared to larger strains than we have been able to consider here.

\section{Summary}
\label{sec:discus}

In this work we have considered a  model of sheared, athermal, frictionless two dimensional spherocylinders in suspension at constant volume.  The simplicity of our model, in which the only interactions are pairwise repulsive elastic forces and a viscous damping with respect to the suspending host medium, allows us to shear to very long total strains and completely characterize the behavior of the system over a wide range of packing fractions $\phi$, strain rates $\dot\gamma$, and particle asphericities $\alpha$.  In two prior works we focused on the rheological properties of  this model and the variation of the jamming transition $\phi_J$ with particle asphericity \cite{MT1}, and on the rotational motion and nematic orientational ordering induced by the shearing \cite{MT2}.  In this work we have focused on  the spatial structure and correlations of the sheared system.  

For a size-bidisperse system of particles, we have considered the average velocity profile to check for shear banding, and we have looked at correlations of the transverse velocity, particle position, the nematic order parameter, and the particle's angular velocity.  We find that, while dense systems near and above jamming can form  shear bands on short strain scales, these bands wander over time and so give rise to the expected linear velocity profile when averaging over long strain scales.
We find that transverse velocity correlations give evidence for a diverging length scale as the jamming transition is approached, however, this is only so for nearly circular particles with small $\alpha=0.01$; for more elongated particles with $\alpha=4$, the location of the minimum in the correlation function seems to decrease to smaller distances as the packing approaches and goes above the jamming $\phi_J$.  We find that the positional and the nematic order parameter correlations remain short ranged, even as the packing $\phi$ approaches and goes above $\phi_J$.  We thus confirm the conclusion of our prior Ref.~\cite{MT2} that the finite nematic order parameter $\mathbf{S}_2$ of the sheared system is not a consequence of long range cooperative behavior among the particles, but is rather because the finite shearing rate $\dot\gamma$ acts like an ordering field.  We also have computed the angular velocity correlation between particles, and find that particles in contact are anti-correlated, while the correlation essentially vanishes at larger distances.  Particles thus rotate incoherently.

For a size-monodisperse system of elongated particles with $\alpha=4$, we have considered several of the same quantities, in order to quantify what structural differences might exist between the monodisperse and bidisperse systems. Considering the velocity profile, as with bidisperse systems we find a similar shear banding on short strain scales that averages to the expected linear velocity profile on long strain scales.  However, unlike the bidisperse system, for dense systems well above jamming we have found that the system can also lock into coherent blocks of particles that move at constant velocity, sliding over one another to give the imposed fixed strain rate.  Measurement of the distribution $\mathcal{P}(\theta)$ of particle orientations also shows a distinct difference from the bidisperse system; whereas in a dense bidisperse system $\mathcal{P}(\theta)$ has a single broad peak, located at a finite angle with respect to the flow direction, in a dense monodisperse system the peak in $\mathcal{P}(\theta)$ stays comparatively sharp and is located close to $\theta=0$, while a shoulder that develops into a secondary peak develops at large $\theta$.  Thus in the monodisperse system the particles are most likely to orient parallel to the flow direction.

We have also computed the positional and nematic order parameter correlations for the monodisperse system and find  a set of sharper peaks that persist to larger distances than in the bidisperse case.  The monodisperse system thus has greater local ordering than the bidisperse system. But still we find that correlations decay exponentially and so correlation lengths remain finite.  Our finding that there are significant structural differences in  dense packings, comparing monodisperse and bidisperse systems, supports our conclusion in Ref.~\cite{MT2}  that there is a crossover from a single-particle-like behavior at small $\phi$, to a behavior dominated by the geometry of the dense packing at large $\phi$, and that this is responsible for the non-monotonic variation we see in the magnitude of the nematic order parameter $S_2$ as $\phi$ increases.  Comparing the magnitude of $S_2$, as well as the average angular velocity $-\langle\dot\theta_i\rangle/\dot\gamma$, for monodisperse vs bidisperse systems, we see that the two are in good agreement  for all $\phi<\phi_{S_2\,\mathrm{max}}$, but then disagree for $\phi>\phi_{S_2\,\mathrm{max}}$.

%Finally we have studied the behavior when we shear starting from  well ordered initial configurations, as compared to the random initial configurations that we use elsewhere in our work.  For both bidisperse and monodisperse systems, we find at small $\phi\lesssim 0.8$ that such configurations get disordered by the shearing and reproduce the same steady-state ensemble as found when starting from an initial random configuration. However for larger $\phi$, particularly in the monodisperse system, the ordering of the initial state can persist to very long strains.  It appears that the stability of such initially ordered states is related to the extent of  misalignment of the particles in the direction transverse to the flow.  

Finally, we have studied the behavior when we shear starting from well ordered initial configurations, as compared to the random initial configurations that we use elsewhere in our work.  In many cases we find that the ordered initial configuration eventually evolves to the same steady-state ensemble obtained from an random initial configuration.  However, it is difficult to generalize about the process that leads to this disordering.  For initial configurations with no particle contacts, particle rotations induced by the viscous drag force lead to the collisions that  are essential to this disordering.  At dilute packings $\phi$, where the free volume available to each particle is larger, we always find that the initial configuration disorders.  But, at denser packings,
whether such collisions are effective to disrupt the particle ordering, or whether they lead merely to the wagging of particles as in Fig.~\ref{mono-wag}, seems to depends on details of the initial configuration.  One factor that increases disordering is when there is greater misalignment of the particle positions $y_i$ in the direction transverse to the flow.  Since the average flow velocity $v_{ix}$ of a particle is set by the particle's coordinate $y_i$, the greater the misalignment of the particles, the more are the collisions that are induced by translational motion; combined with particle rotations, such collisions act to break up the initial ordering of particles in well defined rows.  When there is little vertical misalignment, particles more easily slide over one another, preserving the row ordering.  

The dependence of the time required to disorder on the packing density $\phi$ seems to vary with  the particular initial configuration.  In some cases, such as the phase separated bidisperse configurations of Fig.~\ref{bi-ord-ii}
or the monodisperse configurations with the localized defect of Fig.~\ref{mono-ord-ii}, the configurations seem to disorder faster as the packing $\phi$ increases, (though in both cases the most dense $\phi$ fails to follow this trend).  In other cases, such as the bidisperse configurations of particle stacks in Fig.~\ref{bi-ord-i}  or the monodisperse configurations of Fig.~\ref{mono-ord-i}, disordering takes longer as $\phi$ increases.  We have no clear understanding of why this is so, and  we therefore leave this question for future work.

% Bidisperse - random stacks - all phi disorder quickly, except for the largest phi=0.95
% Bidisperse - phase separated - seems like will disorder except for phi=0.95.  The rate at which they disorder increases with increasing phi; larger phi disorders more rapidly

% Monodisperse - ordered rectangular lattice with random displacements - phi=0.70 disorders quickly, 0.75 takes much longer to disorder, phi=0.80 and larger system does not disorder
% Monodisperse - ordered rectangular lattice with random displacements + defect of rotated particles - all phi disorder except phi=0.95.  Larger phi disorder more quickly.

\section*{Acknowledgements}

This work was supported in part by National Science Foundation Grants No.  CBET-1435861 and No. DMR-1809318. Computations were carried out at the Center for Integrated Research Computing at the University of Rochester. 

%\appendix

%%%%%%%%%%%%%%%%%%%%%%%%%%%

%\clearpage
\bibliographystyle{apsrev4-1}

\end{document}